\documentclass{JHEP3}
\usepackage{axodraw}
\usepackage{epsfig}
\usepackage{rotating}

\newcommand{\rt}[1]{\begin{sideways}#1\end{sideways}}

\newcommand{\twographst}[3]{%
 \unitlength=1.1in
 \begin{picture}(5.8,4.6)(0.5,0.25)
 \put(0,4.84){\epsfig{file=#1, width=0.698 \wth, angle=270}}
 \put(0.9,2.8){\epsfig{file=#12, width=0.68 \wth}}
 \put(2.71,4.84){\epsfig{file=#2, width=0.698 \wth, angle=270}}
 \put(3.61,2.8){\epsfig{file=#22, width=0.68 \wth}}

 \put(0.5,0.2){\epsfig{file=#3, width=0.57 \wth}}
 \put(2.6,0.2){\epsfig{file=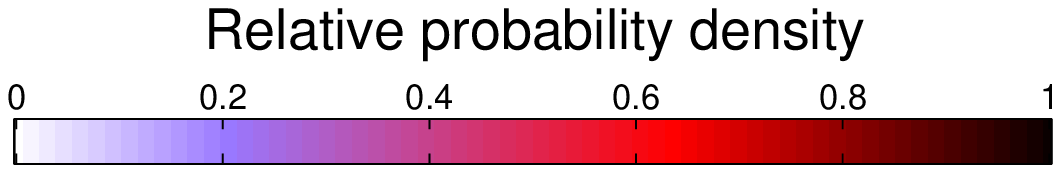, height=1cm, angle=90}}
 \put(3.55,0.2){\epsfig{file=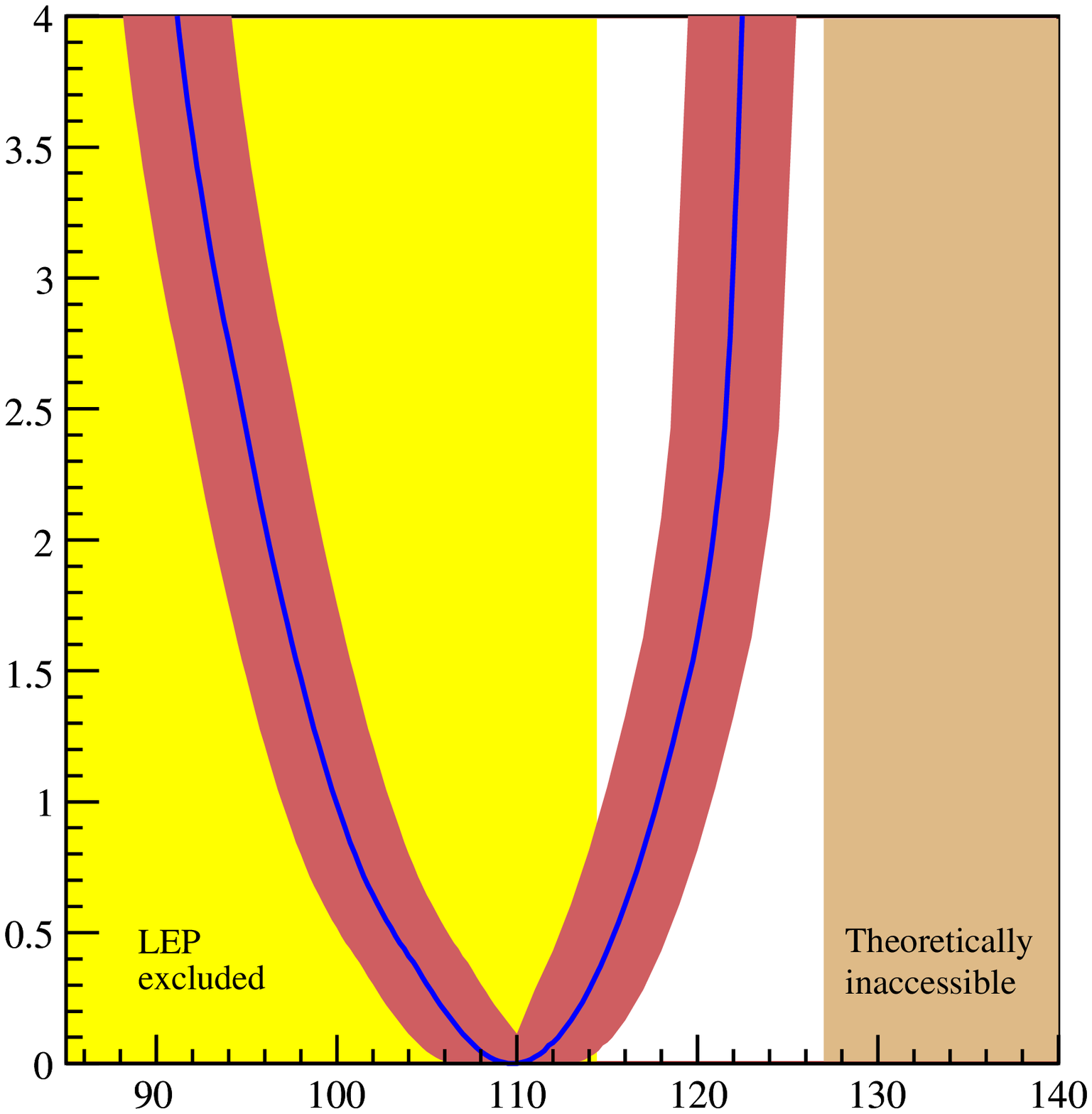, width=0.6 \wth}}
 \put(0.5,4.3){(a)}
 \put(3.2,4.3){(b)}
 \put(0.5,2.3){(c)}
 \put(3.2,2.3){(d)}
 \end{picture}
}

 \newlength{\wth}
\title{SUSY Predictions and SUSY Tools at the LHC}

\author{B.C.~Allanach$^*$, \\
$^*$DAMTP, CMS, University of Cambridge, Wilberforce road, Cambridge, CB3 0WA,
United Kingdom}

\abstract{We provide a bestiary of public codes and other algorithmic tools
  that can be used for analysing supersymmetric phenomenology. We also
  describe 
  the organisation of the different tools and communication between
  them. Tools exist that calculate
  supersymmetric spectra and decay widths, simulate Monte Carlo events as
  well as   those that make
  predictions of dark matter relic density or that predict precision electroweak
  or $b$-observables. Some global fitting tools for use in SUSY phenomenology
  are also presented. 
  In each case, a description and a link to
  the relevant web-site is provided. It is hoped that this review could serve
  as an ``entry-gate'' and map for prospective users. 
}

\begin{document}

\section{Introduction}

Analysis in
high energy particle physics is becoming increasingly complex; the higher
energies and luminosities of current-day colliders lead to higher
multiplicities in events. The current high-energy frontier is dominated
by hadron-hadron colliders, at the Tevatron ($p \bar p$ at 2 TeV) or, in the
near future, the Large Hadron Collider ($pp$ at 10 or 14 TeV), leading to
additional complications in describing hadronic initial states and radiation.
On the theoretical side, the currently most popular solution to
the technical hierarchy problem is the Minimal Supersymmetric Standard Model
(MSSM). A low energy parametrisation of the MSSM contains over 100
parameters. In fact, a truly supersymmetric version of the Standard Model
contains one less parameter than the Standard Model, since the quartic Higgs
coupling becomes a function of the electroweak gauge couplings in the
supersymmetric version, instead of being a free parameter. However, in order
for the MSSM to be phenomenologically viable, supersymmetry (SUSY) must be
broken, and it is in the SUSY breaking sector of the model that the majority
of the free parameters lie. The vast majority of this 100$+$ dimensional
parameter space is ruled out by fairly tight constraints on flavour changing
neutral currents. This is often taken to be evidence of some additional
structure of the model in the flavour sector. High energy boundary conditions
on the supersymmetry parameters that are flavour universal are popular, and
may be motivated by various string (and/or grand-unified theory) models. 

\begin{figure}
\begin{center}
\begin{picture}(200,180)(0,0)
\GBox(30, 20)(170, 80){0.9}
\Line(100,0)(100,150)
\Line(100,150)(0,175)
\Line(100,150)(0,125)
\Line(100,150)(200,125)
\Line(100,150)(200,175)
\SetColor{Black}
\BText(100,0){Detector simulation}
\BText(100,75){Event generator}
\BText(100,55){Matrix element}
\BText(100,35){Parton shower/hadronisation}
\BText(100,100){Decays}
\Line(200,150)(100,150)
\BText(100,150){SUSY Spectrum calculator}
\Line(200,175)(200,125)
\BText(200,150){Global fits}
\SetColor{Red}
\BText(0,175){Theoretical boundary condition}
\BText(0,125){Input observables}
\SetColor{Green}
\BText(200,175){Dark matter}
\BText(200,125){Electroweak/flavour observables}
\end{picture}
\end{center}
\caption{Schematic of the interaction between various
  programs that perform different SUSY phenomenology calculations. The need
  for information exchange is denoted by a line. \label{fig:schem}}
\end{figure}
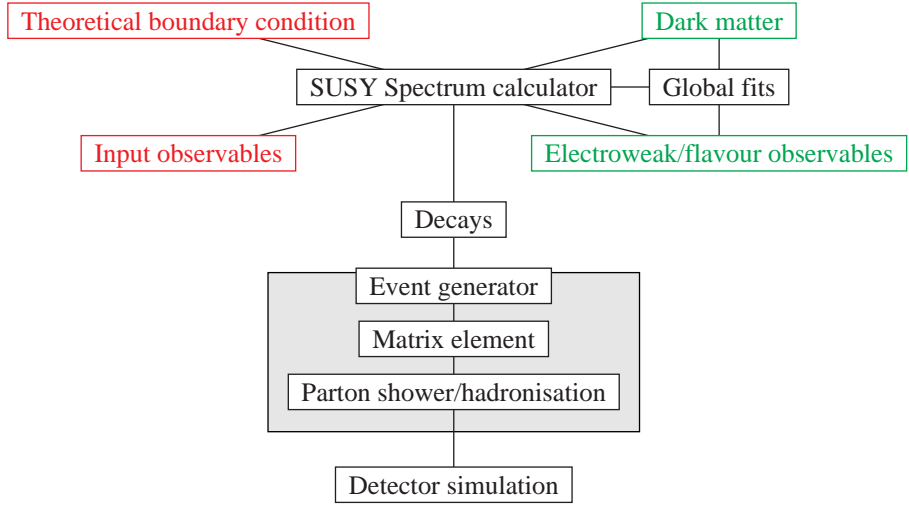
A schematic of SUSY phenomenology calculations is shown in
Fig.~\ref{fig:schem}. Typically, one may want to assume some high energy
theoretical 
boundary condition upon the SUSY breaking sector. One wishes to calculate the
MSSM spectrum and couplings consistent with this boundary condition and some
input observables ($M_Z$, $m_t \ldots$) with a spectrum calculator. The
spectrum and couplings can 
then be passed to another program that calculates decays of the various
sparticles. Once the masses and decays of the sparticles are calculated, this
information may be passed to an event generator in order to randomly simulate
several events in some high energy collision. This process is often split into
two sub-steps: one performing the hard $2 \rightarrow N$ particle collision
(matrix element generation),
and one  
performing hadronic showering, initial state radiation and
decays of the 
sparticles (event generation).  
Experimental colleagues often
then  want to pass such simulated events through a detector simulation in order
to see how the detector might smear the kinematics. 
Alternatively, the spectrum and coupling information could be passed to
packages which calculate indirect observables, such as the dark matter relic
density left in the universe, dark matter direct detection cross-sections, 
electroweak or $b$-observables. These data may then be used in global fits to
the particular SUSY breaking scenario assumed. 
Some of the programs available perform several
of these tasks, but there is currently no single program that performs all of
the 
tasks. Previously, information was passed around on an {\em ad hoc}\/ basis: 
each spectrum generator had to be interfaced separately with each program that
used its output. With $N$ independent codes, the required number of interfaces
such 
that they could each exchange information to the others was $\sim N!$. For
this reason,  
several accords have been written and agreed upon in order to cut down on the
total number of required interfaces, with an associated reduction in the
number of mistakes in the interfacing procedure. 

The SUSY Les Houches Accord (SLHA)~\cite{SLHA} allows information on
the 
masses and decays of SUSY (and some relevant Standard Model) particles to be
passed in between codes. 
The accord is based on ASCII text, in order to allow easy cross-language
communication without introducing platform dependence. The parsing of (files or
memory variables
containing) such ASCII text is an easy task for many human beings, but
the disadvantage of an ASCII format is that developers of tools must write
parsing code. Luckily, even this task has been 
performed, with a SLHA-file parser available~\cite{parser}.
The original SLHA dealt purely with the ``vanilla-MSSM'': inter-generational
sparticle mixing is not taken into account, R-parity and CP are conserved. 
The second SLHA~\cite{SLHA2} generalises the possible MSSM models: R-parity
violating, CP and flavour-violating versions of the MSSM are all specified. In
addition, the most 
popular MSSM extension where a Standard Model singlet chiral superfield is
added, the Next-to-Minimal Supersymmetric 
Standard Model (NMSSM), is covered. 

The original Les Houches Accord (LHA)~\cite{LHA}, allows hard parton-level
events to be passed from matrix element calculators onward down the chain 
to the event generators. It does this by means of a {\tt fortran77} common
block, which specifies properties of the particular process being simulated
such as the
types of particles involved and their momenta. Colour flow in the diagrams
requires particular attention and is encoded in the LHA\@. However, all of the
Les Houches Accords attempt 
to hide such details and requirements from the user. Only tool developers have
to concern themselves with them. More recently, the Les Houches Accord event
record has 
been changed to a minimal {\tt XML}-style structure, for clarity, simpler
parsing and to side-step cross-language difficulties~\cite{LHA2} and several
parsers (in different languages) have been developed and are available.
The accord has also been re-written to take into account potential new beyond
the Standard Model physics models~\cite{LHA3}. 

In this review, I shall briefly describe the publicly available,
supported, documented codes which allow
supersymmetric phenomenological calculations\footnote{In fact, a ``quick
  guide'' of 
SUSY tools was written over two years ago~\protect\cite{Skands:2006sk}. The
present review 
contains an updated and much more extensive overview of the field.}. 
In each case, a link 
to a current web-site and a reference to the relevant 
manual is given. The default language of each program is {\tt fortran77}, but 
if the code is written in a different language, it shall be detailed in this
review at the
point when the main functionality of the code is discussed.
As time passes, it is foreseen that some of the links listed here will become
out of 
date. The reader is advised to read the manual of any code they wish to
use from the electronic {\tt arXiv} web-site in order to find updated links to
downloads etc. In addition, more accuracy and extended functionality will no
doubt be added to the various programs as time passes. This guide is intended
to serve as a snap-shot of documented, supported,  publicly available SUSY
phenomenological tools at the time of
writing. It is not practical to 
continually up-date it as the state-of-the-art evolves. However, it should be
mostly accurate for a good few years and there are plans to extend a
Beyond the Standard Model tools repository~\cite{repos} to a {\tt Wiki site}, 
so that the authors of the codes may include up-dates to the
accuracy/functionality as they occur. 
We shall not describe here any of the detector simulations. 
First, in section~\ref{sec:specdec}, we shall describe codes that calculate
MSSM SUSY spectra and decays. Then, in section~\ref{sec:ME}, we list
matrix element generators, followed by event generators in
section~\ref{sec:events}. We then turn to constraints: in section~\ref{sec:dm}
we review public SUSY dark matter codes, followed by other indirect
constraint calculators in section~\ref{sec:constraints}. 
We review some of the algorithms required to perform global fits to SUSY
models using indirect observables in section~\ref{sec:fits}. Finally, in
section~\ref{sec:summary}, we conclude and present a table summarising
the functionality of the SUSY tools mentioned in this review. 

\section{Spectrum and Decays\label{sec:specdec}}

There are four publicly available dedicated MSSM spectrum generating codes,
displayed in Table~\ref{tab:spects}. They all solve the MSSM renormalisation
group equations (RGEs) to 
two-loop order, subject to two sets of boundary constraints. One set of
boundary constraint is at the weak scale, and matches the MSSM parameters to
current data on Standard Model particle masses and couplings. It also ensures
that electroweak symmetry is broken successfully by adjusting the MSSM $\mu$
parameter. The other boundary condition is typically at a high energy scale,
and involves setting the SUSY breaking parameters according to some
theoretical model of SUSY breaking mediation. Universal mSUGRA, minimal
anomaly mediation and minimal gauge mediation are supported by all of the
codes. In addition, non-universal models such as those that can be invoked by
the SLHA are supported. Each of the codes supports different additional
possible SUSY breaking models. They each also support the SLHA aside from 
{\tt ISAJET}~\cite{ISAJET}. 
An  unofficial version of ISAJET which outputs in SLHA format does exist,
however~\cite{unofficialISAJET}. 

\TABULAR{c|cccc}{Name & Language & RGEs & comment & manual \\ \hline
{\tt ISAJET} & & dominant 3rd& $\nu_R$ & \cite{ISAJET} \\
{\tt SOFTSUSY} & {\tt C++} & 3-family mixing  & & \cite{SOFTSUSY} \\
{\tt SPheno} & {\tt fortran90} & 3-family no-mixing & $\nu_R$ & \cite{SPHENO}
\\ 
{\tt SUSPECT} & & dominant 3rd & & \cite{SUSPECT} \\ \hline
}{SUSY Spectrum generators. $\nu_R$ indicates that the program includes an
  option for including right-handed neutrinos in the spectrum in order to
  obtain neutrino masses. Dominant 3rd RGEs mean that all Yukawa couplings
  other than the third family's are neglected in the RGEs whereas 3-family
  no-mixing 
  means that all diagonal Yukawa couplings are included.\label{tab:spects}}

\EPSFIGURE{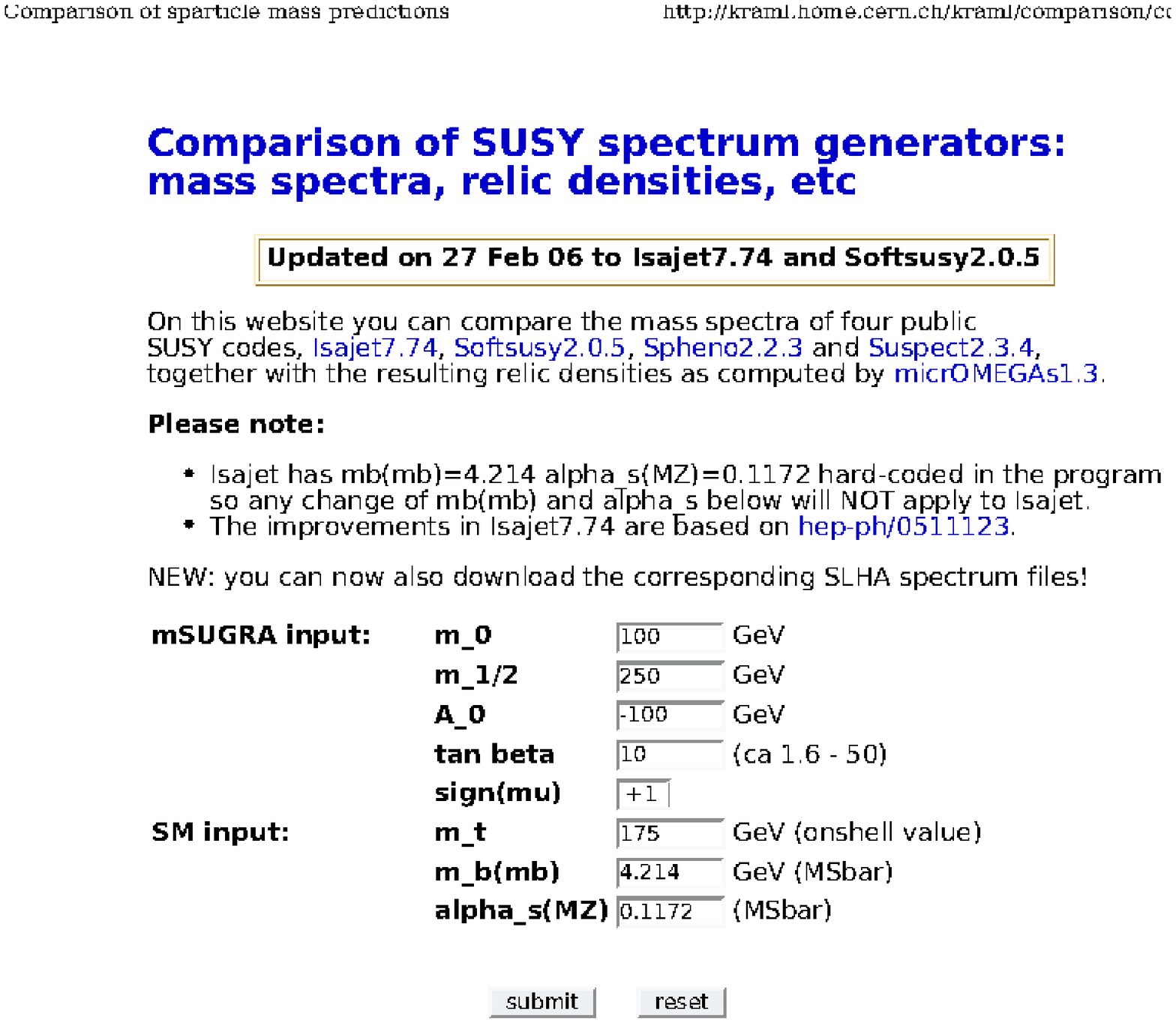,height=8cm}{Comparison web-page mSUGRA form~\cite{sabineWeb} \label{fig:form}}
The details of approximations within the codes can be found within the
manuals, and although similar, do tend to vary somewhat. They may differ by
higher-order corrections, for example. The matching conditions to current data
at the weak scale is mostly in the one-loop approximation. But when one is
correcting a QCD cross section 
with stop loops, for example, in order to extract the MSSM value of
$\alpha_s(M_Z)$ that should be used, the question arises which stop mass
should be used? The codes either use pole masses or running masses evaluated
at different scales for the stops running in the loops. The difference
between these choices is actually a higher 
order effect, and could only be fixed were the two-loop matching conditions
known.  
Broadly speaking, RGE evolution is two-loop, 
in different approximations, as displayed in Table~\ref{tab:spects}.
{\tt ISAJET}~differs from the other codes in that it decouples sparticles at
their mass scales, thus re-summing terms $\sim {\mathcal O}(1 / (16 \pi^2)\log
[ \Delta M / M])$, 
where $\Delta M$ is the splitting between two sparticles and $M$ is their
average mass. On the other hand, the other three codes all evolve using MSSM
RGEs above 
$M_Z$ without decoupling sparticles, but then one-loop decoupling effects are
added to the weak-scale boundary condition to leading logarithmic order. This
latter approach allows the easy addition of some one-loop finite pieces, some
of which are missed by the mass-scale decoupling approach taken by {\tt
  ISAJET}. In summary, one may 
expect the mass-scale decoupling approach to provide a more accurate answer
when sparticle 
splittings are very large, and $M_Z$ decoupling including all finite terms to
be more accurate for a more 
typical sparticle splitting. The codes all agree to the percent level, except
in particularly difficult parts of parameter space such as the focus point or
very large $\tan \beta$~\cite{uncertainties}, where the differences can be
much larger due to inherent numerical instabilities and the size of higher
order corrections in those regions. 
Fortunately, a web-site exists~\cite{sabineWeb} where one can input a SUSY
breaking point on a web form as exemplified in Fig.~\ref{fig:form}, and
quickly compare the output from the different codes. If 
one is doing a study on a particular point, for example, this provides a quick
practical way of finding out if the point comes with particularly large
theoretical uncertainties or not.  

\subsection{MSSM Sparticle Decays \label{sec:spdecay}}

Currently, the programs that calculate sparticle decay branching ratios are
{\tt Herwig++}~\cite{HERWIG} ({\tt C++}), {\tt ISAJET}~\cite{ISAJET}, {\tt
  MadGraph}~\cite{MADGRAPH}, 
{\tt
  PYTHIA}~\cite{PYTHIA}, {\tt SDECAY}~\cite{SDECAY} and  
{\tt SPheno}~\cite{SPHENO}.  {\tt SDECAY}, {\tt PYTHIA} and {\tt Herwig++} take the SLHA stream from any of
the other codes in order to produce SLHA-compliant output including decay
information, whereas the other two codes are linked to their spectrum
generators. The decay packages implement tree-level
two-body decays of fermions and gauginos and 
three-body decays of charginos, neutralinos and gluinos.
{\tt SPheno} includes gluonic QCD corrections into decays by quarks.
{\tt SDECAY}
implements some three and four-body decays of top squarks and one-loop
corrections to the two-body decays.
{\tt PYTHIA} and {\tt Herwig++}
contain internal routines for calculating sparticle decays, including
tri-linear $R-$parity breaking effects.  
{\tt Herwig++} and {\tt MadGraph} include {\em angular correlations}\/ between
subsequent decays 
in a sparticle cascade decay using the pioneering techniques of
ref.~\cite{Richardson:2001df}, whereas all the other codes make a
phase-space  
approximation. The program {\tt
  BRIDGE}~\cite{BRIDGE} was written in order to decay particles 
passed to it by matrix element generators in general models defined in the
{\tt MadGraph} format, then pass them on to
showering and hadronisation programs. It calculates two and three-body
tree-level decays
itself, while keeping track of initial vertex spin structures via {\tt HELAS}.
Typically, phase-space is a reasonable approximation in
hadronic collisions unless one
is trying to fit the spin of sparticles. 

\subsection{Higgs Masses and Decays}

There are some packages specialising in SUSY Higgs calculations: {\tt
  FEYNHIGGS}~\cite{FEYNHIGGS} calculates the Higgs masses in a Feynman
diagrammatic approach. In calculating Higgs masses,
important two-loop effects are included for the MSSM with or without complex
parameters, with a 
re-summation of the leading (s)bottom corrections. One-loop non-minimal
flavour violating corrections to Higgs masses/mixings are included at the
one-loop level. 
The program calculates 
the Higgs spectrum and decays 
and provides an estimate of 
theoretical uncertainties in the prediction. The two-body tree level decays
include dominant one-loop corrections and the Higgs decays to $gg$ and $\gamma
\gamma$ 
include all of the MSSM particles in the loop. 
A {\tt FEYNHIGGS} web-form interface exists for
checking single points in parameter space~\cite{FEYNHIGGS}. 
The program {\tt CPsuperH}~\cite{CPSUPERH} also
performs MSSM Higgs calculations when CP violating phases are present,
including some effects up to two-loop order. 
The program is based on
renormalisation-group improved diagrammatic calculations that include
logarithmic as well as threshold corrections and $b$-quark Yukawa coupling
re-summation.
Some
dominant one-loop pieces are included in the Higgs decays, which can be into
SUSY or SM particles (including some important three body decays). The Higgs
couplings and neutral Higgs mixings are also provided by the program. 
{\tt HDECAY}~\cite{HDECAY} calculates up to three-loop QCD corrected decays of
Higgs 
bosons in the CP-conserving MSSM where expressions exist in the
literature, including some loop-induced decays, 
decays into two massive gauge bosons, 
three-body decays and decays into SUSY particles. Leading electroweak
corrections are included (they can become important in the large Higgs mass
regime due to enhanced Higgs self-interactions). All MSSM particles are
included in the loop 
for the calculation of $\gamma \gamma$ and $gg$ Higgs decay modes. The leading QCD corrections are included for the gluonic mode. 
{\tt FCHDECAY}~\cite{FCHDECAY} computes the flavour changing neutral current
decays $BR(H^0 \rightarrow {\bar t}c,t {\bar c})$ and $BR(H^0 \rightarrow
{\bar b}s,b{\bar s})$ in the flavour violating MSSM, 
using SLHA2 for input/output. It includes full one-loop SUSY QCD
contributions. 

\subsection{NMSSM}

The addition of a Standard Model singlet superfield to the MSSM constitutes a 
potential solution to the $\mu$ problem (why $\mu$ is of order the electroweak
scale rather than some much heavier scale) and is called the
next-to-MSSM (NMSSM). In the package {\tt NMSSMtools}~\cite{NMSSMTOOLS},
sparticle masses are calculated using two-loop NMSSM
RGEs in the dominant third family approximation. Tree level sparticle decay
widths and branching ratios
are also calculated. 
The Higgs masses, couplings and widths (for two-body modes) are calculated
within the NMSSM using approximations to the one and two-loop dominant
corrections.  For decays into the SM particles,
the widths are calculated including one-loop SM QCD
corrections.

\section{Matrix Element Generators and Cross Section
  Calculators\label{sec:ME}} 

In the high-energy LHC r\'{e}gime, often we wish to calculate the production
of more than two hard particles. This is the job of matrix element 
generators. Matrix element generators can usually calculate total or
differential cross-sections and/or produce independently sampled
events. 
Simulating (for example) the production of two
squarks plus some additional hard QCD radiated jets requires us to deal with 
 complicated Feynman diagrams involving many particles in the final state. 
For this job, one uses a matrix element generator, which simulates or
calculates the hard
process (e.g.\ $q {\bar q} \rightarrow {\tilde t}_1 {\bar {\tilde t}}_1 +
jet$). The matrix element generators are currently mostly at tree-level,
particularly as regards SUSY physics. In practice, $2 \rightarrow 6$ to $2
\rightarrow 8$ processes
may be feasible depending upon the number of Feynman diagrams, although a vast
amount of CPU time may be needed to 
compute them (using, for example, the grid). The number of Feynman diagrams
tends to grow to be too large with increasing numbers of final-state
particles.

{\tt FeynArts}/{\tt FormCalc}~\cite{feynarts} are Mathematica  packages for the
generation and 
calculation of Feynman diagrams up to one-loop order. They can thus be used to
calculate matrix elements for scattering
processes.  Up to $2 \rightarrow 3$ processes can be calculated at the one-loop
level with integration optimisation, although {\tt FormCalc} has been
successfully used to compute $2 \rightarrow 4$ processes at tree-level. 
Vanilla, CP violating and non-minimal
flavour violating versions of the MSSM are available. There is also a way of
encoding some new physics Lagrangian model for extensions. {\tt FormCalc}
simplifies the amplitudes generated by {\tt FeynArts}
 analytically and generates 
{\tt fortran77} code for the numerical evaluation of the squared matrix
element. Automatic generation and pictorial representation of Feynman diagrams
is also supported, as is convolution with parton density functions (PDFs).
Recently, a program {\tt HadCalc}~\cite{HADCALC} has been
developed based on {\tt FeynArts} and {\tt FormCalc}. It takes the output from
those codes in terms of partonic cross sections and convolutes them with
PDFs. There are convenient ways to place cuts and an interactive menu-driven
front-end that can be used to dial in SUSY parameters. 

Currently, {\tt CalcHEP}~\cite{CALCHEP} and {\tt CompHEP}~\cite{COMPHEP} ({\tt
  C})\footnote{These two programs have the same origin, but 
at some stage the development of them branched. Because of this, although the
programs are now different, many features of them are similar.} can cope with
up to 6  
external legs in a Feynman diagram, for example ($1 \rightarrow 5$ or $2
\rightarrow 4$). The two programs can produce {\tt C} output of analytical
expressions for 
subsequent compilation and use. They have graphical interfaces which can display
modulus squared Feynman diagrams. Models are already defined for the MSSM,
NMSSM and the CP-violating MSSM\@. For the encoding of model Lagrangians and
parameters, {\tt LanHEP}~\cite{LANHEP} ({\tt C}) is used. If the user wishes
to extend some SUSY model outside of the ones already defined, 
{\tt LanHEP} provides the means.
{\tt MadGraph}~\cite{MADGRAPH}
performs vanilla MSSM matrix element
calculation with SLHA input. Helicity amplitudes are constructed based on the
{\tt HELAS}~\cite{HELAS} library in order to encode spin information of the
produced particles, which can be used in their decay. Feynman diagrams are
drawn, and {\tt fortran77} output is produced for the matrix element.
A Monte-Carlo integrator package has been included in {\tt MadGraph}, and SUSY
differential or total cross sections can be calculated using it.
Alternatively, the final result can be Les Houches Accord formatted
parton-level events that
can be passed into an event generator for subsequent parton showering and
hadronisation. The {\tt MadGraph} web-site has a form 
that can be filled in to get events returned automatically. The event
generator {\tt SHERPA}~\cite{SHERPA} utilises an event generator {\tt
  Amegic++}~\cite{AMEGIC} ({\tt C++}) that also uses the helicity amplitude
technique and calculates at the tree-level, with the possibility of up to six
particles in the final state of the hard scattering.
{\tt
  Whizard}~\cite{WHIZARD} ({\tt fortran95}) includes initial and final state
polarisations and 
can calculate in the vanilla MSSM as well as the CP-violating case. It uses
{\tt O'Mega}~\cite{OMEGA} ({\tt 
  O'Caml}) to
translate a helicity amplitude into computer code as needed. {\tt O'Mega} is
designed with special tricks to avoid the factorial increase in CPU time with
the number of external particles. 
It has been demonstrated to work for
some processes with eight particles in the final state. {\tt
  SUSYGEN}~\cite{SUSYGEN} is restricted to $2 \rightarrow 2$ SUSY production
processes. It
can include polarisation in $e^+ e^-$ collisions and
covers vanilla as well as R-parity or CP-violating MSSM models. 
{\tt GRACE}~\cite{GRACE} performs computations of $e^+e^- \rightarrow$ up to
four bodies in the MSSM at tree-level. {\tt GRACE} draws the relevant Feynman
diagrams for the user.

Numerical results of several hundred SUSY production cross-sections were
compared between
{\tt MadGraph}~\cite{MADGRAPH}, {\tt SHERPA}~\cite{SHERPA}
and {\tt
  Whizard}~\cite{WHIZARD}  and they were all found
(eventually) to 
agree~\cite{comparison2}. 
{\tt PROSPINO}~\cite{PROSPINO} ({\tt fortran90}) computes MSSM next-to-leading
order cross sections for the 
production of two sparticles at hadron colliders. It can also cope with the
production of weak gauginos in the split SUSY framework. 
Detailed calculations of cross-sections of $e^+e^- \rightarrow$ sleptons at
the one-loop level are also available from {\tt ILCslepton}~\cite{FREITAS}.
{\tt FeynHiggs}~\cite{FEYNHIGGS} calculates Higgs production cross-sections
for the Tevatron and the LHC including SUSY corrections at the production
vertex. 

\section{Event Generators\label{sec:events}}

\EPSFIGURE{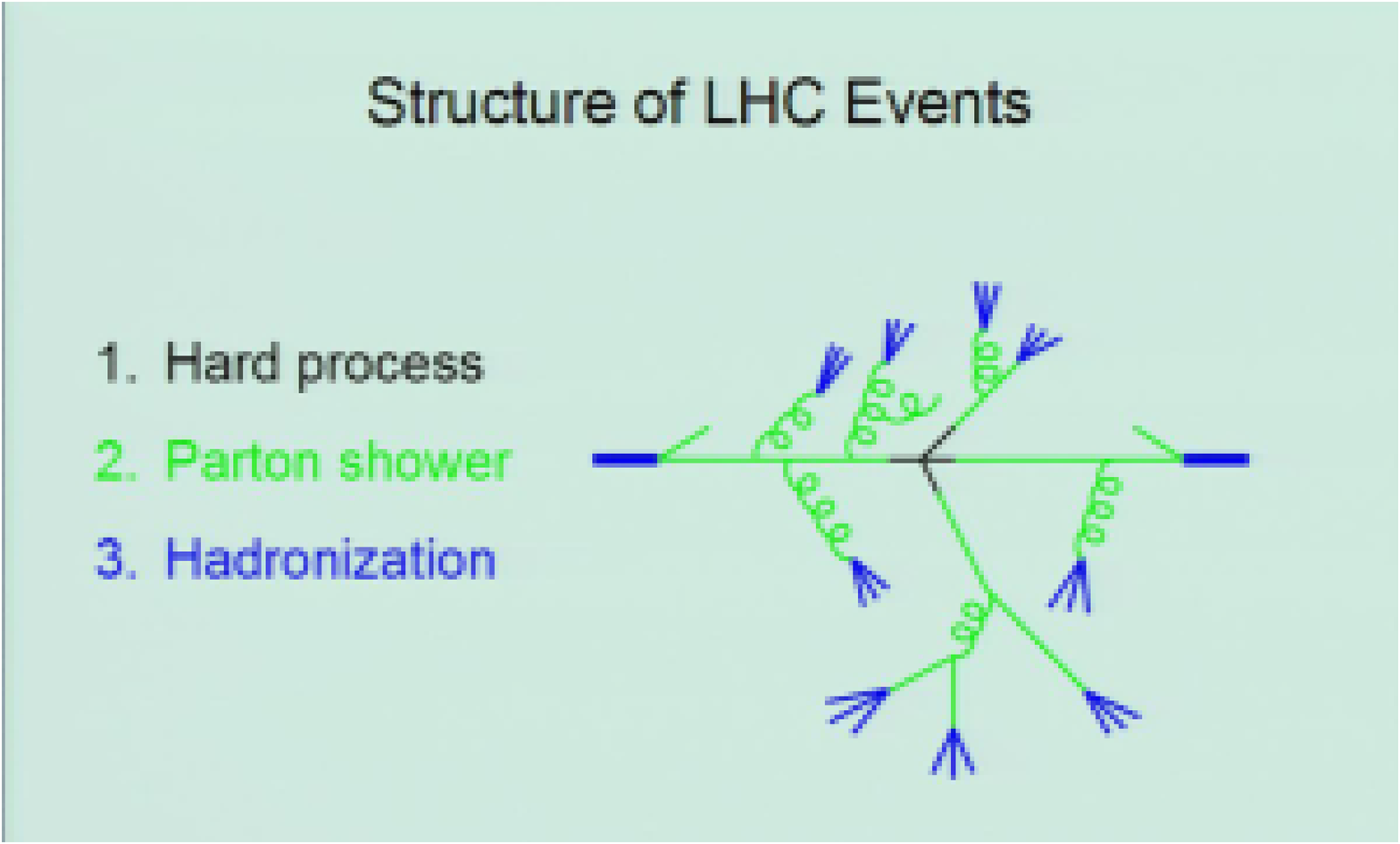,height=6cm}{Schematic of a hadron
  collision simulation in an event generator. \label{fig:structLHC}} 
The most well-known general
purpose SUSY event generators ({\tt PYTHIA}~\cite{PYTHIA}, {\tt
  Herwig++}~\cite{HERWIG}) usually implement  
a hard-sub process in terms of two particles scattering on two particles,
represented by the central vertex in Fig.~\ref{fig:structLHC}.
The initial particles in this hard-sub process may be leptons, or point-like
constituents of hadrons. In the latter case, the quarks and gluons are
extracted from a hadron by means of the parton density functions.
The hard-sub process is usually calculated at leading order in perturbation
theory within event generators, especially for exotic signals such as SUSY\@.
If SUSY particles are produced in the simulated event, they are then decayed
randomly, according to the branching ratios calculated by a program in
section~\ref{sec:spdecay}. The resulting cascade decay spits out SM particles,
some of which may be quarks or gluons. These quarks and gluons then emit soft
QCD radiation, which is modelled by the parton shower. Parton showering
encodes the fact that the matrix elements of massless coloured particles
emitting a gluon have a singularity in the infra-red or collinear limit. The
initial state may also shower, emitting QCD radiation. It can be important to
include effects in the shower coming from colour coherence in order to
describe the resulting jets adequately. 
Various properties such as angular ordering of the shower (with preceding
emissions being at smaller angles) are evident in the resulting event. 
Once the partons are showered down to some energy scale to be decided by the
event generator, some non-perturbative modelling (and a tune to data) collects 
the partons together into hadrons which, after their decays have been
simulated, may be observed in the
detector. Finally, the simulated events are often represented by a series of
lines in a text file (representing variables held in memory), each line
describing the kinematics and state of a particle involved with the event. 
Information on which particles decayed into which other particles is
also indicated in this event record. We briefly mention some of the
Standard Model properties of event generators, since elements of them are also
relevant to SUSY events, but the interested reader should see
ref.~\cite{Dobbs:2004qw} for a more complete 
guide to Standard Model event generators in hadron collisions.

A general framework for encoding new physics models is included in {\tt
  Herwig++}~\cite{HERWIG} so that users may define the relevant particles and
Feynman rules
for the hard sub-process. The {\tt MSSM} has already
been defined within the framework, but extended SUSY models must be input by
using it.
It can read SLHA files for
input SUSY information.
The {\tt Herwig++} shower algorithm treats
QCD radiation from coloured heavy objects (for example tops). It can evolve to
zero transverse momentum of emissions, giving an improved simulation of the
dead-cone effect for radiation from massive particles. An eikonal multiple
scattering model is used for simulating additional partonic collisions in the
same hadronic collision. Such processes form part of the underlying event. 
{\tt Herwig++} uses a cluster model for the hadronisation step, clustering
quarks and gluons that have similar kinematics into colour singlet states,
which decay to hadrons and hadron resonances. {\tt Herwig++} treats its decays
including spin correlations all of the way down the various decay chains. 
{\tt Herwig++} relies on an underlying {\tt C++}
structure developed for high energy collisions called {\tt
  ThePEG}~\cite{peg}. 

{\tt PYTHIA}~\cite{PYTHIA}, on the other hand, does not depend upon {\tt
  ThePEG}, but is 
stand-alone.  As well as hadron-hadron collisions, it can deal with $e^+ e^-$
beams. Initial and final-state parton showers are based on 
$p_T$-ordered evolution, terminating at 1 GeV. Although there is currently a
{\tt C++} version in 
development, it does not contain any SUSY physics and so we concentrate on the
older {\tt fortran77} version in this review. Many different options for the
changing the models of parts of the {\tt PYTHIA} simulation are possible, but
here we describe the default models.
Hadronisation and hadron fragmentation (decay)
are modelled by the Lund string model, where hadrons are modelled
to be a colour flux tube, ended where the (di-)quarks are located. 
The MSSM, the NMSSM, tri-linear $R-$parity violation, as well as long-lived
coloured sparticles such as those that exist in models of split SUSY, are
included in the {\tt PYTHIA} distribution. Polarisation is included for $e^+
e^-$ incoming 
beams. All decays of sparticles are using the phase-space approximation, and
so sparticle spin is not simulated. 

A new event generator has recently been developed called {\tt
  SHERPA}~\cite{SHERPA} ({\tt C++}). The main design feature of {\tt SHERPA} is
that it combines parton shower evolution and matrix element generation. The
MSSM with or without CP violation and full inter-generational mixing is
included, as well as a general formalism (compatible with the 
matrix element generator {\tt Amegic++}~\cite{AMEGIC}, which ships with the
{\tt SHERPA} distribution) provided for adding
new particles and interactions. {\tt Amegic++} can practically handle up to
six jets in the 
final state for $e^+ e^-$ collisions, and up to three jets for hadron
collisions.  
One of the difficulties of combining parton showers and matrix element
generation for hard jets is the problem of double-counting. 
If one simply adds
the matrix element generation for 3 jets to the 2-jet plus parton shower
sample, one could easily double count the region where  
one of the jets is soft (and therefore already included in the parton shower).
In {\tt SHERPA}, the parton shower evolution and matrix element
generation are matched via the CKKW formalism~\cite{CKKW}, 
where the matrix element configurations are re-weighted according
to a pseudo shower history and shower emissions that overlap with higher
order matrix elements are rejected.
{\tt SHERPA}
also performs the hadronisation/fragmentation step using its own cluster
model~\cite{SherpaCLModel}, which includes di-quark spin effects and a dynamic
separation of the r\'{e}gimes of clusters and hadrons according to their
masses and flavours. 
When calculating SUSY decays in {\tt SHERPA}, there is currently no facility
for picking the decay products automatically, the user must supply which decay
chain is required. After this has been done though, the spin information and
off-shell effects are included in each sparticles' decay into the next
sparticle and Standard Model particle. 

The {\tt ISAJET}~\cite{ISAJET} event generator simulates $pp$, $p {\bar p}$
and $e^+ e^-$ collisions at high energy, based on perturbative QCD and
phenomenological models for parton and beam jet fragmentation, not including
colour coherence effects: the probability of emitting a soft gluon is
multiplied by a factor given by the Alterelli-Parisi function. {\tt ISAJET}
keeps only the parts that are in the exact collinear limit, but uses
non-collinear kinematics. 
QCD radiation from initial and final states is
simulated. Sparticle pair production at tree-level is supported, along with
subsequent decay. The {\tt ISAJET} hadronisation model is the independent
fragmentation ansatz of Field and Feynman, which forms new (di-)quark-anti-quark
pairs out of partons, and groups them together into mesons and baryons with
some fraction of their summed momenta. 

\section{Predictions of SUSY Dark Matter\label{sec:dm}}

The recent WMAP5 cosmological fits to the cosmic microwave background and
other data provide us with an accurate observation of the density of dark
matter in the universe as a fraction of the relic density: $\Omega h^2=0.1143
\pm 0034$~\cite{wmap5}. The MSSM offers several possible candidates for 
weakly interacting massive particles that could play the r\^{o}le of cold dark
matter, since the lightest supersymmetric particle (LSP) is stable from the
assumption of $R-$parity. The dark matter candidate obviously must be without
electric charge, so that it does not interact with light, and also should be
colourless, otherwise it would have fused with nuclei during nucleosynthesis
and been discovered in anomalously heavy isotope searches. 
Gravitinos and lightest neutralinos are possible candidates within the MSSM,
although in extended models, other SUSY particles are possible dark matter
candidates.
In principle, SUSY dark matter may be discovered in direct detection
experiments, where nuclear recoils from collisions with SUSY dark matter are
possible. If the dark matter candidate interacts too weakly (for example in the
case of the gravitino), the direct detection cross-sections are far too small
to small to be seen in the foreseeable future. However, in the case of the
neutralino, there is a chance for direct dark matter observation. 
In order to detect dark matter on earth, it is of course a necessary
condition that there is some dark matter going through the detector. While
only a small amount is empirically known about the small-scale structure of
dark matter halos, numerical $N$-body simulations indicate that even if one
starts with strict filaments or cusps of dark matter, subsequent Newtonian
evolution
will tend to smear it out. Thus, the prospect of having our galaxy's dark
matter localised completely elsewhere in the galaxy seems unlikely, but 
it should be borne in mind that any calculation of the local dark matter flux
on earth is subject to large astrophysical modelling uncertainties. 
Aside from the direct detection of particulate dark matter, there are
prospects for indirect detection, where for example, dark matter annihilation
in the sun, in the earth or in the centre of the galaxy produces high energy
particles 
that can be detected on earth or on satellites. 

Once a SUSY model's parameters has been fixed and a cosmological model is
assumed, it is possible to estimate the amount of current dark matter relic
density in the universe is predicted. The codes tend to assume the $\Lambda$
CDM model of a  
cosmological constant plus cold dark matter component, since this is quite a
simple and good fit to the WMAP and large scale structure data. 
One has to track the abundances of the different species of
sparticle through the early evolution of the universe. They can annihilate
with each other into Standard Model particles, 
but eventually, the expansion of the
universe makes the sparticles too far apart to interact. 
Aside from losses due to
annihilation processes, each sparticle will end up as a LSP through its decays. 
The tracking of the abundances of the various sparticle species and 
involves the evolution of coupled Boltzmann equations. There are many
different annihilation cross-section processes to consider, and the relevant public
codes currently calculate at tree-level. The velocity distributions of the
SUSY particles are derived from Maxwell-Boltzmann approximations. 
This can still involve the
calculation of thousands of Feynman diagrams, however. As a consequence, the
current tools calculate mainly at tree-level. However, loop corrections can
give large $\sim O(10)\%$ effects in some cases~\cite{sloops}.

{\tt DarkSUSY}~\cite{DARKSUSY} contains hard-coded matrix elements for the
many different annihilation processes of the vanilla MSSM\@. It can calculate
the relic density as well 
as direct and indirect detection rates, with a choice of different nuclear
form factors for the direct rates. Solar system WIMP velocity distributions
can be used to calculate the capture in the Earth of dark matter particles. 
An exotic component of positron, anti-proton and anti-deuteron in cosmic rays
originating from neutralino pair annihilation in the galactic halo can be
calculated. Hadronisation and fragmentation was calculated with {\tt PYTHIA}
and the results tabulated from various neutralino masses, which {\tt DarkSUSY}
interpolates in order to provide an estimate of the particle yield. 
For particle yields coming from annihilation in the earth and the sun, 6
fundamental channels are included: $c {\bar c}, t {\bar t}, b {\bar b}, \tau^+
\tau^-, W^+W^-$ and $Z^0Z^0$. Recent solar and terrestrial
density models are included as a necessary ingredient in the calculation.
For galactic halo annihilations, 
$W^+W^-, Z^0Z^0,
W^+H^-, Z^0h^0, 
Z^0H^0$, $h^0A^0$ and $H^0A^0$ channels are included, with subsequent
decay of these particles, including the
heavy quarks $c, b$ and $t$. 
The $gg$, $\gamma \gamma$ and $Z \gamma$ channels
occurring at the one 
loop level are also included. 
Anti-matter production yields from
dark matter annihilation in the Galactic halo are determined by {\tt DarkSUSY}.
Modelling the propagation of anti-matter is
non-trivial, but {\tt DarkSUSY} attempts this through various approximations
which can be found in the manual.
High
energy neutrinos and neutrino-induced muons can be detected by neutrino
telescopes and their yields are calculated. 
The SLHA is currently not
supported, but instead dedicated interfaces to {\tt ISAJET} and {\tt SUSPECT}
are included for spectrum generation. 
There is a web interface linked from the {\tt DarkSUSY} homepage for inputting a
MSSM model and calculating the relic 
density and detection cross-sections. Thus, if one is doing an analysis on one
point in parameter space, one can check its dark matter properties easily
on-line.  

{\tt IsaRED} is part of the {\tt ISAJET}~\cite{ISAJET} package, and calculates
the relic density of neutralino dark matter in the MSSM\@. Annihilations between
$\chi_1^0$, 
$\chi_2^0$, $\chi_1^\pm$, ${\tilde e}_1$, ${\tilde \mu}_1$, ${\tilde \tau}_1$, 
${\tilde \nu}_e$, ${\tilde \nu}_\mu$, ${\tilde \nu}_\tau$, ${\tilde u}_1$,
${\tilde c}_1$, ${\tilde t}_1$, ${\tilde d}_1$, ${\tilde s}_1$, ${\tilde b}_1$
and gluinos are taken into account in the calculation. In the same package, 
{\tt IsaRES} evaluates spin-independent and spin-dependent direct detection
rates. Squark, $Z^0$ and Higgs exchanges are included at tree-level and 
neutralino-gluon interactions involving quarks, squarks and Higgs bosons are
included at the one-loop level. 

{\tt micrOMEGAs} ({\tt C})~\cite{MICROMEGAS} calculates the relic density of the
LSP at tree-level and
direct/indirect detection 
rates in the    
vanilla MSSM, the MSSM with complex phases and the NMSSM\@. Important
higher-order QCD and SUSY QCD corrections to Higgs quark vertices are included.
The program can be used to calculate the relic density of a charged and/or
coloured next-to-lightest supersymmetric particle. This can be useful in the
case of a gravitino LSP\@. Gravitinos are not simulated by {\tt micrOMEGAs}, but
a simple formula can be used to extract their relic density from the relic
density of the next-to-lightest supersymmetric particle. The most important
annihilation channels for any given model point can be output. 
{\tt micrOMEGAs} uses {\tt
  CalcHEP}~\cite{CALCHEP} in order to calculate any necessary Feynman
diagrams, and so 
extensions can be encoded using the {\tt LanHEP}~\cite{LANHEP} Lagrangian
formulation, for models where there is only one stable particle. 
Only diagrams that may contribute up to some specified fraction (by default,
$10^{-5}$) of the
thermally averaged total annihilation cross-section are included, which makes
for faster computation.
LSP scattering rates on nucleons and nuclei in the 
spin-independent and spin-dependent interaction cases are also presented.
$\gamma$, $e^+$, ${\bar p}$ and $\nu$ yields for indirect
direction purposes (at $v \rightarrow 0$ and/or in the continuum) are
calculated.   
Like {\tt DarkSUSY}, {\tt micrOMEGAs} uses the basic channels $q {\bar q}$,
$\mu^+ \mu^-$, $\tau^+ \tau^-$, $W^+ W^-$ and $Z^0 Z^0$ and interpolates
tables of $\gamma, e^+, {\bar p}$ and $\nu$ production as obtained by {\tt
  PYTHIA}. For channels that contain two different particle species $AB$, the
final 
spectrum is obtained by taking the average of di-$A$ production and di-$B$
production as a rough approximation. The galactic gamma-ray flux is
calculated with a modified 
isothermal distribution of dark matter in the galaxy. For direct detection
rates, higher order corrections to Higgs-quark vertices and one-loop 
neutralino-gluon interactions are included
for the vanilla MSSM, the CP-violating MSSM and the NMSSM.

\section{Predictions for constraints\label{sec:constraints}}

There are many constraints upon supersymmetric models: direct constraints tend
to be the easiest to implement, being (usually) phrased as lower bounds on
sparticle 
masses. Relevant indirect constraints are upon branching ratios for rare
decays, precision electroweak observables or electric dipole moments for
example, and often occur at the loop level. Being of general utility,
{\tt FormCalc}~\cite{feynarts} can be used to calculate the relevant SUSY matrix
elements.

\subsection{$b$ observables}

The branching ratio of $b \rightarrow s \gamma$ has long been used to
constrain supersymmetric models, and is calculated by several codes. 
Many of the codes calculate it in the vanilla MSSM without SUSY flavour mixing.
For minimal flavour violating MSSM computations, {\tt SusyBSG}~\cite{SUSYBSG}
calculates the branching ratio for the decay $b \rightarrow s \gamma$
taking into account all of 
the available next-to-leading order (NLO) contributions, 
including the complete supersymmetric two-loop QCD corrections 
to the Wilson coefficients of the magnetic and chromo-magnetic operators, as
well as 
an improved NLO determination of the relation between the 
Wilson coefficients and the branching ratio. 
{\tt micrOMEGAs}~\cite{MICROMEGAS}, predicts the branching ratio
including next-to-leading order contributions for the Standard Model. The
charged Higgs and supersymmetric large $\tan \beta$ effects beyond
leading-order are included. 
{\tt
DarkSUSY}~\cite{DARKSUSY}  performs
a NLO calculation which is complete for the
Standard Model prediction and adds some dominant NLO MSSM corrections. 
{\tt SPheno}~\cite{SPHENO} and {\tt SUSPECT}~\cite{SUSPECT} include one-loop
MSSM corrections and some NLO 
corrections to the branching ratio. 
{\tt SuperIso}~\cite{SUPERISO} calculates the $b \rightarrow s \gamma$
branching ratio in the vanilla MSSM with flavor violation, NLO
supersymmetric contributions and next-to-next-to-leading order (NNLO) Standard
Model contributions. Flavour violation is supported through the SLHA2
interface. 
{\tt SuperIso} is currently the only code to predict the
isospin symmetry breaking $\Delta_{0-}$ of the $B \rightarrow K^* \gamma$
decay including the NLO SUSY contributions. 
{\tt CPsuperH}~\cite{CPSUPERH} can provide a prediction for the branching
ratio as well as as its CP-asymmetry and SUSY contributions to
$B_{s,d}^0-{\bar B}_{s,d}^0$ mass differences ($\Delta M_{s,d}$).
{\tt FeynHiggs}~\cite{FEYNHIGGS} provides a prediction for the $b \rightarrow
s \gamma$ branching ratio including non-minimal flavour violating effects. 
Another $b-$physics observable that can constrain SUSY is the to-date
unobserved rare decay mode $B_s \rightarrow \mu^+ \mu^-$. 
The SUSY calculation in {\tt micrOMEGAS}~\cite{MICROMEGAS} includes the one-loop
contributions due to chargino, sneutrino, stop and Higgs exchange. $m_b$
re-summation effects at high $\tan \beta$ are taken into account. 
{\tt CPsuperH}~\cite{CPSUPERH} also performs the calculation of $BR(B_s
\rightarrow \mu^+ \mu^-)$ in the CP
violating MSSM, as well as $B_d \rightarrow \tau^+ \tau^-$, $B_u \rightarrow
\tau^+ \nu_\tau$. Each branching ratios is calculated in the single-Higgs
insertion approximation.  
{\tt NMSSMtools}~\cite{NMSSMTOOLS} calculates $b \rightarrow s \gamma$,
$B_s\rightarrow \mu^+ \mu^-$, and $B^+ \rightarrow \tau^+ \nu_\tau$ branching
ratios as well as $\Delta M_{s,d}$  in the NMSSM at one-loop order. 
{\tt ISATOOLS}~\cite{ISAJET} includes NLO contributions to some of the Standard
Model Wilson coefficients for $BR(b \rightarrow s \gamma)$ and one-loop MSSM
corrections. Branching ratios for $B_s \rightarrow \mu^+ \mu^-$ and $B_d
\rightarrow \tau^+ \tau^-$ are calculated to one-loop, using approximations
for the chargino masses (neglecting their mixing). 
The fitting program {\tt SuperBayes}~\cite{SUPERBAYES} uses the {\tt
  micrOMEGAs} prediction of $BR(b \rightarrow s \gamma)$ at NLO and then
augments it by NNLO Standard Model QCD 
contributions.

\subsection{Anomalous magnetic moment of the muon}

The anomalous magnetic moment of the muon is currently around 3$\sigma$ higher
than the Standard Model prediction. There is thus room for a non-zero SUSY
contribution. {\tt ISATOOLS}~\cite{ISAJET}, {\tt SPheno}~\cite{SPHENO}, {\tt
  SuperIso}~\cite{SUPERISO}, 
{\tt
  micrOMEGAS}~\cite{MICROMEGAS} and {\tt 
  DarkSUSY}~\cite{DARKSUSY} calculate 
the predicted SUSY contribution to one-loop order, whereas {\tt
  FEYNHIGGS}~\cite{FEYNHIGGS} and {\tt SUSPECT}~\cite{SUSPECT} also include
some two-loop corrections.  

\subsection{Electric dipole moments}

For calculations of electric dipole moments in the CP-violating MSSM, {\tt
  micrOMEGAS}~\cite{MICROMEGAS} can provide estimates for the electron and
Thalluim. One-loop 
neutralino/chargino contributions and two-loop squark, quark and chargino
contributions are included as well as four-fermion operators for Thallium. 
Two-loop Higgs-mediated contributions to electron, muon and Thallium electric
dipole moments are calculated in {\tt CPsuperH}~\cite{CPSUPERH}. However,
currently some 
well-known one-loop contributions have yet to be implemented.
The Thallium, neutron and mercury electric dipole moments are calculated in {\tt
  FEYNHIGGS}~\cite{FEYNHIGGS}. 

\subsection{Electroweak observables}

{\tt micrOMEGAs}~\cite{MICROMEGAS} and {\tt SUSPECT}~\cite{SUSPECT} can output
the $\Delta \rho$ parameter, which describes some
loop corrections to electroweak observables. They both contain one-loop
stop/sbottom 
contributions, as well as two-loop QCD corrections due to gluon exchange and
the heavy-gluino limit of gluino exchange. {\tt FeynHiggs}~\cite{FEYNHIGGS}
also contains a calculation of $\Delta \rho$, with corrections up to two-loops. 
{\tt SPheno}~\cite{SPHENO} outputs the one-loop sfermion contributions to
$\Delta \rho$. 
In terms of the electroweak observables themselves, {\tt FeynHiggs} also
computes $M_W$ and $\sin^2 \theta_w^{eff}$ including some two-loop SUSY
contributions, non-minimal flavour
violating effects and the effect of complex phases in the stop/sbottom sector
at one-loop. 

\section{Fitting tools\label{sec:fits}}

We first introduce some necessary statistical terms, then go on to discuss
their use in the context of SUSY fits. 
Typically, global fits of models to data utilise a statistical ``figure of
merit'' for each point in parameter space to characterise how well it fits
data. The most familiar one for particle physicists is probably $\chi^2$, 
but sometimes likelihood is used instead. 
Likelihood $L$ can be simply related to
the $\chi^2$ parameter, $L \propto e^{-\chi^2/2}$. $L$ or $\chi^2$ are often
quoted in frequentist statistical interpretations of data.
Bayesian statistics
turn these quantities into probability distributions on the input parameters
of the model, requiring the introduction of the infamous prior probability
distribution. The probability distribution of some parameter {\em after}\/
confrontation with data is called the posterior probability distribution. 
A global fit of some model to data often consists of finding the variation of
the figure of merit with the model parameters. The best-fit set of model
parameters is sometimes
quoted, with the amount of parameter space contained within some expected
amount of statistical variation of data. More complete analyses map out the
figure of merit on the parameter space, and Bayesian analyses then make
probabilistic inferences based upon the map. 

\subsection{Algorithms for multi-dimensional fits\label{sec:alg}}

Even if one restricts the MSSM to some lower number-of-parameters form such as
mSUGRA, the parameter space is still of considerable dimensionality: 4 for
a given sign of $\mu$ ($m_0, M_{1/2}, A_0$ and $\tan \beta$). Also, if one wants
to perform global fits of the model to 
data, one should include variations of the relevant Standard Model input
parameters. $m_t$ is proportional to the largest parameter in the model, the
top Yukawa coupling, and for high $\tan \beta$ the bottom Yukawa coupling, 
proportional to the bottom quark mass, can change the predicted values of
observables. Variations of $\alpha_s$ 
within its empirical uncertainties can also have a large effect on squark and
gluino masses through the RGE evolution, since it is the largest gauge
coupling. If one is including precision electroweak observables in the fit,
including uncertainties on the fine structure constant $\alpha$  becomes
essential. Thus, in mSUGRA one has an eight-dimensional relevant parameter
space.  
Scans in such a space are impractical, since the required number of points 
is exponential in the number of parameters. If one required a resolution of 25
points for each parameter, 1.5$\times 10^{11}$ points would be required in
total. To make matters worse, there are often sharp features in the $\chi^2$
distribution that would render such a low resolution insufficient. Such a
large number of points cannot be calculated in a reasonable amount of CPU
time, even given recent advances in computer technology. If one has access to
a computer farm, calculating a few million points is feasible within a few
days, for example 
(unless one wants to simulate event generation, which would take much
longer). There is therefore a need for more sophisticated scanning algorithms
that can reduce the required number of scanned points for parameter spaces of
more than three dimensions.

The software tool {\tt MINUIT}~\cite{MINUIT} is a well-tried function
minimiser. It calculates derivatives of the figure of merit with respect to
input parameters and performs
hill-climbing algorithms to try to find the best-fit point. It then determines
the error matrix from a matrix of second derivatives of $\chi^2$.
This error matrix contains information about the $1\sigma$ standard
deviations of the parameters in the Gaussian
approximation (where $a\chi^2$ is assumed to be parabolic around the 
best-fit point) and including correlations. For cases where the Gaussian approximation is a bad
approximation, another internal {\tt MINUIT} algorithm can be used for
determining errors including 
non-linearities, but can be very time consuming depending upon the amount of
non-linearity. Algorithms that use derivatives can be problematic when the
surface that they are minimising are rough. In the SUSY fitting case, the
original SUSY spectrum is obtained by an iterative process up to some numerical
accuracy, which then feeds into the rest of the figure of merit calculation,
providing small discontinuities in the surface. 
A typical numerical fractional accuracy in this stage of the calculation might
be $10^{-3}$. 
While a fractional accuracy of $10^{-5}$ is feasible, it
requires much more CPU time per 
scanned point, and is actually unattainable in certain ``difficult'' regions
of parameter space such as the focus-point region. {\tt MINUIT} also finds 
parameter degeneracies problematic, where the figure of merit does not change
much along some curve in parameter space. Despite these short-comings, {\tt MINUIT} has been used to perform global fits of mSUGRA to
global 
data successfully~\cite{Buchmueller:2007zk}. 

MCMC methods are commonly used in
cosmological~\cite{roberto,COSMOMC} and other 
contexts, and recently 
there has been a realisation that they are very useful to the SUSY
high-dimensional scanning problem. MCMCs scan more often where the fit is good
and 
the figure of merit is high and less often in the tails of distributions. 
In fact, the density of scanning is proportional to the figure of merit. 
MCMC methods have a high CPU overhead, meaning that they are not the most
efficient tool for one or two dimensional problems. But the required number of
points goes roughly linearly with the number of dimensions rather than
exponentially, and so they are very useful for our higher-dimensional mSUGRA
fitting problem. 
In this context, a Markov chain consists of a long list, or ``chain'' of
points and their associated likelihoods. Statistical inference can be made by 
binning these points in terms of some quantity of interest.
The simplest implementation of MCMC is called the Metropolis
algorithm~\cite{metropolis}. In the Metropolis algorithm, for the first point
in the chain, a point $x_0$ is picked at 
random in parameter space and its posterior density calculated, $p_0$. 
A potential next point $x_1$ is picked in the vicinity of the previous point,
again at random. If $p_1 > p_0$, the new point is accepted. Otherwise, the new
point is accepted with probability $p_1 / p_0$. If the new point is {\em not}
\/accepted, the previous point is added {\em again}\/ on to the chain. This
algorithm is repeated many times, until it has explored all
of the relevant parameter space. There are many choices of how to pick a
potential next point ``in the vicinity'' of the current one, and some trial
and error is usually involved in setting the length scales involved. Usually,
a Gaussian function is used to randomly choose the distance of the new point
away from 
the current one, but formally, any well behaved function would work in the
limit of an infinite number of MCMC steps provided it has no true zeroes. 
For efficient scanning, the
length scale should be of order the length scale of the likelihood
variation. If it is much larger, hardly any new points will be accepted and
the efficiency will be too low. If it is much smaller, many new points will be
required to explore all of the good-fit parameter space. In order to verify
that the algorithm has indeed explored the parameter space properly, it is
good practice to run several statistically independent chains concurrently. 
One can then compare the results in the different chains statistically to see
how similar they are~\cite{gelmanRubin}. The Metropolis algorithm does not
rely on derivatives and is therefore immune to serious problems caused  by
roughness from numerical error. It can easily be used to interpret data in a
Bayesian form or in 
a frequentist form. For the Bayesian inference, one plots the quantities in
question (say, squark mass vs gluino mass) in bins. The marginal posterior
probability distribution in terms of these parameters is then proportional to
the number of points in the chain that land in each bin. {\em Marginal}\/ refers
to the fact that all other parameters have been integrated over. 
In order to interpret the chain in a frequentist fashion, one plots the
profile likelihood: the likelihood of the maximum likelihood point that lands
in each bin~\cite{Allanach}. Such a procedure, provided a sufficient number of
samples to get 
near the maximum for each bin has been obtained, is equivalent to minimising
$\chi^2$ in each 
bin. Confidence limits can be found in the parameter plane in question by
plotting iso-$\Delta \chi^2$ contours, where $\Delta \chi^2$ is $\chi^2$
assigned to each bin minus the $\chi^2$ of the minimum bin. MCMC methods thus
provide full maps of the figure of merit across parameter space or other
scalar quantities that one is interested in. A package {\tt
  SuperBayeS}~\cite{SUPERBAYES} ({\tt fortran77}, {\tt fortran90} and {\tt
  C++}) is available for performing global fits to SUSY models using MCMC and
{\tt SOFTSUSY}~\cite{SOFTSUSY}, {\tt DarkSUSY}~\cite{DARKSUSY} and {\tt
  FEYNHIGGS}~\cite{FEYNHIGGS}. The MCMC routines were adapted from {\tt
  cosmomc}~\cite{COSMOMC}, as well as some of the plotting routines. 
The program {\tt SFITTER}~\cite{SFITTER} is currently
being developed which will fit SUSY models to collider data on sparticle masses 
using MCMC methods. 

A problem that is not addressed by either {\tt MINUIT} or by the Metropolis
algorithm is that of well-separated $\chi^2$ minima. {\tt MINUIT} only finds a
local minima. In principle, the Metropolis algorithm may find all local
$\chi^2$ minima in the limit of infinite number of samples. In practice
however, if the local minima are small and require small length scales for 
suggesting proposed points, and the distance in parameter space between them
is large, the chance to ``hop'' from one local minimum to the other may be 
tiny and require an unfeasibly large number of samples. A ``tweak'' to the
Metropolis algorithm exists which can solve this problem and is called bank
sampling~\cite{bank}. In bank sampling, one performs a two-step process. In
the first step, many different Metropolis chains are started and run for a small
number of steps, but numerous enough to find points somewhere near local
likelihood maxima. These points then form the ``bank'' or ``cache'' of points 
used in a new modified Metropolis algorithm. On each MCMC step, there is a
small probability that the chain will propose a point in the vicinity of one of
the bank points. If the new point is added successfully to the chain, the
chain ``teleports'' to the other local maxima. In this way, the relevant local
maxima all appear in the fit results, correctly normalised with respect to
each other.

If only the global likelihood maximum, or equivalently, the global $\chi^2$
minimum, is desired, a different modified Metropolis algorithm
called {\em simulated annealing}\/ can be used. 
Simulated annealed can be used to find a point near a  global
$\chi^2$ minimum when several local ones exist. 
In simulated annealing, it is imagined that
the $\chi^2$ surface is some potential energy surface upon which a particle
moves. A finite temperature is set, which increases
the length scale of the proposal step (or, in the analogy, the average
distance the particle moves). 
The temperature
is very large at the start of the algorithm and gradually decreases to one 
thereafter. The chance of acceptance of a worse-fit point is also fixed to be
higher with increased temperature $T$, being set to $e^{-\Delta \chi^2/T}$,
where $\Delta \chi^2$ is the $\chi^2$ difference between the current point and
the proposed worst-fit one.
In the early stages, the algorithm is more likely to traverse bad-fit
regions and not be trapped in local minima. 
The computer code {\tt FITTINO}~\cite{FITTINO}
can fit a 24-parameter simplified weak-scale MSSM to assumed cross-section and
mass 
from SUSY signal collider data. Tree-level values of observables and subsets
of SUSY 
parameters are used to obtain start values for the $\chi^2$-fit.  
Simulated annealing is then performed in order to find a better approximation
to the global $\chi^2$ minimum. Using these parameter values, {\tt MINUIT} 
is performed in order to minimise $\chi^2$ more precisely. In order to
investigate the uncertainties in the fit, a series of fits for many imagined
experimental data are performed in {\tt FITTINO}, with data smeared around
their nominal values, and the global $\chi^2$ minimum is found in each case. 

In frequentist statistics, hypothesis testing often reduces to finding the
minimum $\chi^2$ of different models. However, in Bayesian statistics, one
wishes to calculate the evidence ratio: the ratio of 
volumes under the posterior probability surfaces, 
a quantity that can be very computationally intensive to calculate. 
Bank sampling provides a method for the rough computation of the evidence
ratio,  
by having bank points within each of the separate models. After the MCMC has
run, the ratio of points in each model is an estimate of the evidence ratio.
Such an estimate may not be very accurate, particularly where the evidence
ratio is much larger or smaller than one. In such cases, one can artificially
multiply one of the model's likelihoods by a factor which will bring the 
resulting evidence ratio closer to one. The normalisation can be un-done, with
the result that the ratio can be computed with smaller 
statistical uncertainty from the likelihood re-scaling. 
The disadvantage of bank sampling for Bayesian
evidence evaluation is that only ratios of the evidence can be determined, not
the evidence value on its own. 

An 
algorithm which solves this problem as well as the well-separated likelihood
maxima problem 
in a completely different way is the `MultiNest'
technique~\cite{Feroz:2007kg}. MultiNest models the multi-dimensional 
likelihood surface with a series of (possibly overlapping)
ellipsoids. Clustering algorithms are contained within the larger
algorithm. They determine when 
an ellipsoid is to be broken up into two different ellipsoids because the
initial 
one does not model the underlying distribution well enough. 
Many live points are chosen, sampled from the prior probability distribution. 
Current live points are described in terms of ellipsoids, determined by
the covariance matrix of the live points, enlarged by about 20$\%$ to take
non-linearities into account.
The live point with the smallest likelihood is replaced with one with a higher
likelihood re-sampled
from the ellipsoids.
Thus, the live points gradually home in on the likelihood maxima as the
algorithm proceeds and the evidence can be calculated from the list of live
points and their evidences, as can posterior probability inferences. 
The evidence of a single model can be accurately calculated in this approach,
in contrast to the case of bank sampling. Thus, as one builds up a list of
different models that one is testing against some set of data, there is no
need to run many different comparisons between the different pairs of models:
a single computation for each model suffices. The nested approach does need to
be able to sample efficiently from the prior probability distribution and so
will not work efficiently in cases where there is no analytic form for the
prior. For extremely high dimensional cases (say, 10 and above), a MCMC-hybrid
nested sampling approach may be more efficient than the ellipsoidal
approach~\cite{Feroz:2007kg} for multi-modal distributions. 

\subsection{Example Global Fits to mSUGRA}

We now display some example results using the various techniques introduced in
section~\ref{sec:alg}. We pick examples of global fits to mSUGRA in the
literature as an example. Typically, the data that authors have chosen to fit
to include the relic
density of dark matter to the relic density of neutralinos, $M_W$, $\sin^2
\theta_w^{eff}$, $BR(b \rightarrow s \gamma)$, $BR(B_s \rightarrow \mu^+
\mu^-)$, $(g-2)_\mu$ $m_t$, $m_b(m_b)$, $\alpha_s(M_Z)$, $\alpha(M_Z)$ and
direct exclusion limits from colliders. 
\FIGURE[t]{\twographst{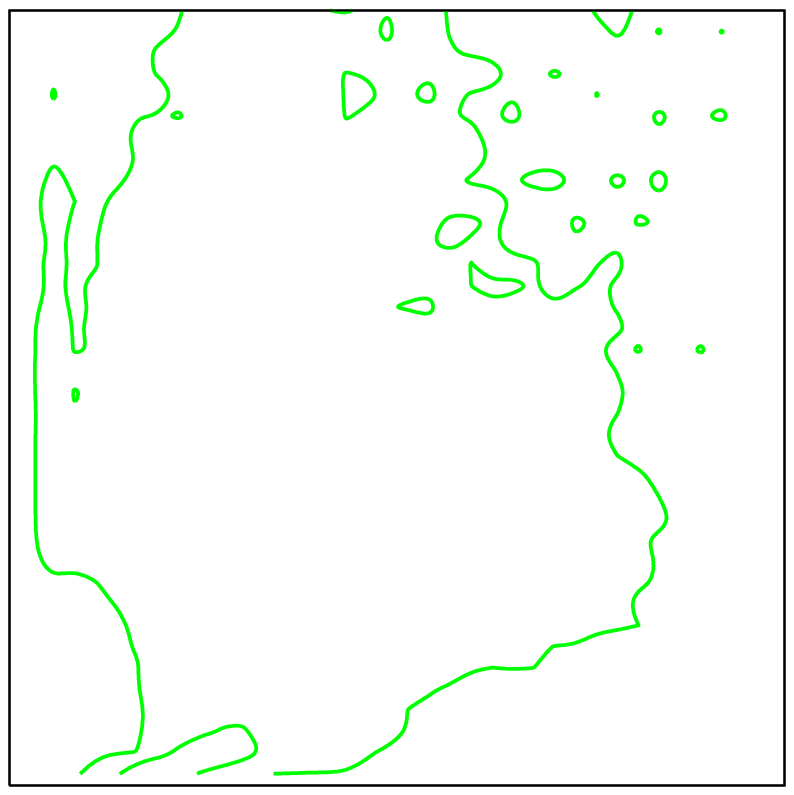}{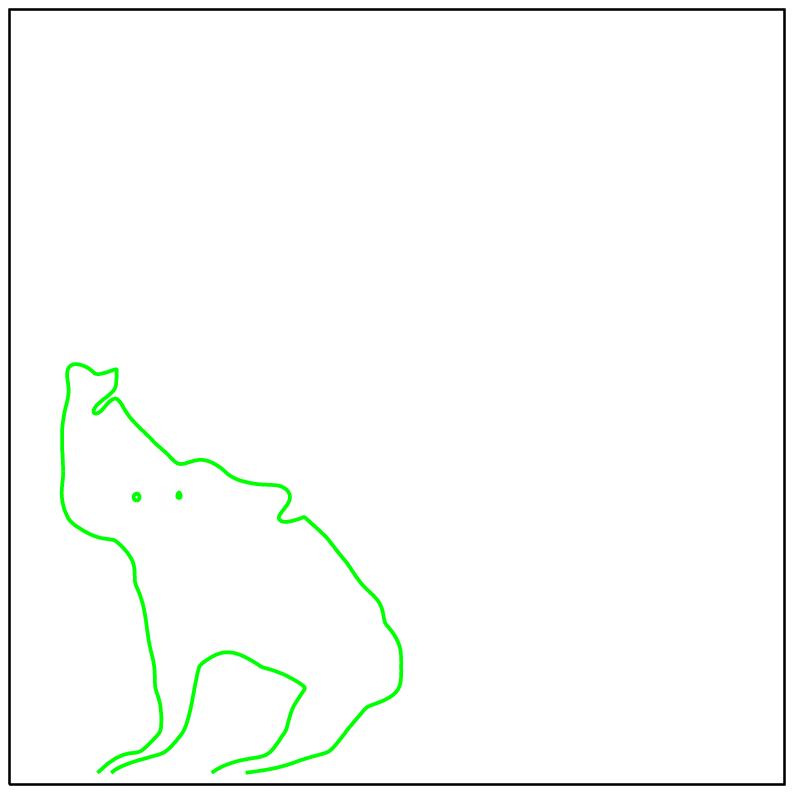}{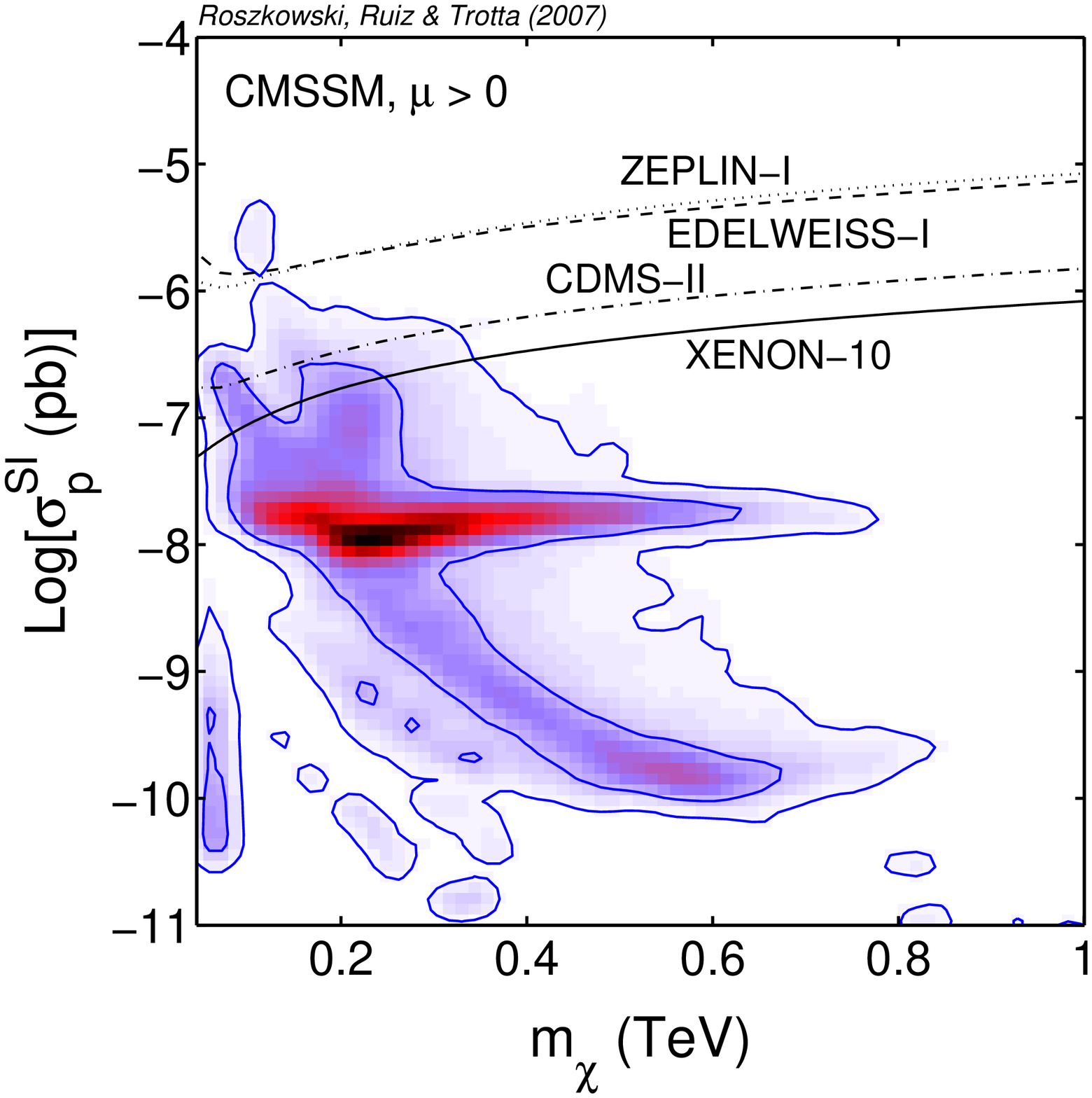}{}
\caption{Global mSUGRA fits in the $m_0-M_{1/2}$ plane: (a) shows the
  Bayesian 
  posterior probability distribution~\cite{Allanach}, (b) shows the frequentist 
  interpretation in the same plane~\cite{Allanach}, (c) displays the direct spin independent
  detection cross 
  section posterior probability distribution function versus mass of the
  lightest neutralino for $\mu>0$ along with some 95$\%$ C.L. exclusion
  contours from    direct   detection experiments~\cite{Roszkowski:2007fd}. 
 Inner and outer contours show the 68$\%$ and 95$\%$
  confidence level regions respectively. (d) shows the $\Delta \chi^2$ of the
  lightest CP-even Higgs 
  mass from all constraints except for the direct LEP2 Higgs mass
  constraint~\cite{Buchmueller:2007zk}.\label{fig:m0m12}}}  
Fig.~\ref{fig:m0m12}a shows the posterior probability distribution in terms of
the $m_0-M_{1/2}$ plane for both signs of $\mu$ with flat priors in $m_0$,
$M_{1/2}$,   $\tan \beta$ and $A_0$ after an MCMC fit using bank
sampling~\cite{Allanach}. The probability relative to the one in the
maximum-posterior bin is shown by the colour, and measured by the bar
on the right. The most probable region at low values of $m_0$ and $M_{1/2}$
corresponds to the stau co-annihilation region, where the lightest stau and
lightest neutralino are quasi-mass degenerate. The extended probability mass
at $m_0, M_{1/2} \sim 0.5$ TeV corresponds to the $A^0$ boson resonance at
high values of $\tan \beta$,
where dark matter annihilation proceeds efficiently through $\chi_1^0 \chi_1^0
\rightarrow A^0 \rightarrow b {\bar b}$. At larger values of $m_0$, we have
the focus point region, where efficient annihilation into weak gauge boson
pairs is possible. In Fig.~\ref{fig:m0m12}b, the same data is interpreted in a
frequentist fashion using the profile likelihood technique. This technique
picks out the best-fit points, rather than averaging over all points in the
unseen dimensions. Figs.~\ref{fig:m0m12}a, b differ where there are
significant volume effects, that is where the volume of points in the unseen
dimensions enhances or diminishes the Bayesian fit. The fact that the
frequentist 
interpretation differs from the Bayesian one can be seen as a signal that more
data is required for the fit; indeed we should not be surprised since a
complex model with eight free 
parameters has been fit with some fairly indirect data. 
Similar fits to the Bayesian ones above were performed
using the Metropolis MCMC algorithm, resulting in quite similar posterior
probability densities for the particle physics properties, despite some
differences in the indirect constraints
used~\cite{Roszkowski:2007fd}. In addition, ref.~\cite{Roszkowski:2007fd}
constrains dark matter detection cross-sections. We show the posterior
probability 
distribution function of the spin-independent direct dark matter detection
cross-section in Fig.~\ref{fig:m0m12}c for flat priors in $m_0$,
$M_{1/2}$,   $\tan \beta$ and $A_0$, for $\mu>0$. The most constraining direct
detection experiment, XENON-10, can be seen to cover some of the favoured
region already, assuming that the flux of dark matter passing through
XENON-10 is the same as the galactic average. In Fig.~\ref{fig:m0m12}d,
we see 
the results of a more 
traditional frequentist $\chi^2$ mSUGRA fit using {\tt MINUIT} in terms of the
lightest CP-even 
Higgs mass of the MSSM~\cite{Buchmueller:2007zk}. For each value of the
lightest Higgs mass, a $\chi^2$ minimisation was performed against all of the
other mSUGRA parameters. Many additional electroweak and $b$ observables were
included in the $\chi^2$ of this fit, although for comparative purposes the
LEP2 direct bound on the Higgs mass was left out. This bound is plotted as the
yellow excluded region in Fig.~\ref{fig:m0m12}d. It can be seen that the global
$\chi^2$ minimum occurs at $m_h \approx$110 GeV, below the direct 95$\%$
C.L. lower bound of 114.4 
GeV. The authors of Ref.~\cite{Buchmueller:2007zk} use the $\chi^2$ curve to 
infer that $m_h=110^{+8}_{-10} \pm 3 $ GeV, where the first uncertainty is
statistical and the second uncertainty is theoretical.

\subsection{Data archival}

The samples from MCMC fits take some effort and CPU time to obtain. 
In principle, the mSUGRA fits could be useful to other physicists, who wish to
make their 
own inferences about observables. If a new calculation of SUSY contributions
to some observable were to come on-line, statistical inference could be made
by simply obtaining some independent samples from the MCMC chains and
calculating the new observable for each. A few thousand points might suffice
in terms of statistics. Then, the totality of current empirical knowledge
about SUSY corrections to the observable is obtained without a need for
complicated multi-dimensional fitting procedures. 
Aside from that, other physicists might be interested in using the chains for
their own scans over the points in parameter space that are compatible with
current data. 
For this reason, the authors of Ref.~\cite{Allanach} have formed the {\tt
  KISMET} web-site, which contains links to text files of the
chain data. 
The weight of each point (the number of times it was visited in the MCMC
procedure), along with the values of input parameters, resulting
indirect observables, sparticle masses and the likelihood, are listed in the
files. 
Also, 10 000 independent samplings from the chains in SLHA format are
available from the web-site. 

The {\tt SuperBayeS}~\cite{SUPERBAYES} data are now available to some extent
on-line: one can fill out a web-form in order to automatically receive plots
of the posterior probability density in dimensions specified by the
user~\cite{SPLOTTING}. These dimensions can be specified to be observables,
input parameters or sparticle masses. 

\section{Summary\label{sec:summary}}

Currently, little is empirically known about supersymmetry except for a few
indirect data. This tends to lead to under-constrained SUSY models and
consequent degeneracies in global fits. However, if the LHC provides some
signals that 
are compatible with SUSY, aside from being an extremely exciting discovery
beyond the Standard Model, hypothesis tests against alternative models and
even between different classes of SUSY models will be desirable. 
Ideally, constraints upon the SUSY Lagrangian would be derived with the help
of high energy $e^+e^-$ linear collider experiments. 
A well defined theoretical framework is needed when higher order corrections
are included in trying to reconstruct a fundamental SUSY theory and its
breaking mechanism. For this purpose, the Supersymmetry Parameter Analysis
(SPA)~\cite{SPA} scheme provides a consistent set of conventions and input
parameters, as 
well as a repository for programs which connect parameters in
different schemes and relate the Lagrangian parameters to quantities that may
be more directly extracted from physical observables
such as masses,
mixings, decay widths and production cross sections for supersymmetric
particles. 

There is a somewhat bewildering proliferation of computer program tools for
SUSY calculations and phenomenology in the literature. This proliferation is a
useful 
development, 
and reflects the interest of the high energy physics community
in supersymmetry. Even more useful is the collusion, collaboration and
organisation between the different programs, to allow results from one to be
fed into another program and interpreted automatically. 
The SUSY Les Houches Accord is a
good example of such practice, and it has now become essential for any
relevant computer tool to use it so that it can communicate with the other
tools. The most popular computer languages that the tools are written in are
still various versions of {\tt fortran} and {\tt C++}. Following the move of
most new high energy physics experiments to {\tt C++}, there is a tendency for
new event generators to be written in {\tt C++} rather than {\tt fortran}.
In fact the
precise language a SUSY tool is written in is becoming less important with the
advent of the communication accords, which are in {\tt ASCII} format. 
Many of the more sophisticated matrix element or event generators use their
own encoding of a Lagrangian to enable the user to define new models. This
approach is of obvious use and generality, but packages have many different
definitions of the 
Lagrangian. Perhaps there is a need for yet another
accord, so that the same model can easily be fed in to different 
tools without the need for the user to translate the Lagrangian between the
various different conventions. 

Finally, we end with a brief summary of the phenomenological SUSY tools that
are covered in this review. Since some programs have several different
functions, we summarise them all together in Table~\ref{tab:sum}, although the
programs are loosely grouped according to their functionality. 
\begin{sidewaystable}
%\begin{table}
  \tiny
  \centering 
  \begin{tabular}{|c|c|cccccc|ccccccccc|ccccccccccc|c|c|c|} \hline
Tool & \rt{${\mathcal L}$} &
\rt{Spectrum} &
\rt{$\nu_R$} &
\rt{RPV} &
\rt{NMSSM} &
\rt{CPV} &
\rt{FV} &
\rt{Decays} &
\rt{Decay Spin} &
\rt{ME} &
\rt{Initial pol} &
\rt{$\sigma_{SUSY}$} &
\rt{$e^+e^-$} &
\rt{$pp$ } &
\rt{Events} &
\rt{PS/Had} &
\rt{$\Omega_{DM} h^2$} &
\rt{$\sigma_{SI, SD}^p$} &
\rt{Ind. DM} &
\rt{$b \rightarrow s \gamma$} &
\rt{$B_s \rightarrow \mu^+ \mu^-$} &
\rt{$\Delta_{0-}$} &
\rt{$\Delta M_{B_s}$} &
\rt{$B \rightarrow \tau \nu, \tau \tau$} &
\rt{$(g-2)_\mu$} &
\rt{EW} &
\rt{edm} &
\rt{Fits} & 
\rt{Web form} &
\rt{code}
\\ \hline
{\tt NMSSMtools}~\cite{NMSSMTOOLS} & %L
& $\surd$ %spec
&  %nu_R
& %RPV
& $\surd$ %NMSSM
& %CPV
&  %FV
& $\surd$ %Decays
& %Decay spin
& % ME
& % ME spin
& % sigma_SUSY
&  % e+e-
& % pp
&  % Events
&  % PS/Had
&  % DM
&  % SI,SD
&  % Ind DM
& $\surd$ % b->sg
&  $\surd$ % Bs->mumu
&  %Delta_0-
& $\surd$ % Delta MBs
& $\surd$ % b->tau nu
& % (g-2)_mu
& % EW
&  %edm
&  & & \\ % fits 
{\tt SOFTSUSY}~\cite{SOFTSUSY} & %L
& $\surd$ %spec
&  %nu_R
& %RPV
&  %NMSSM
& %CPV
& $\surd$ %FV
&  %Decays
& %Decay spin
& % ME
& % ME spin
& % sigma_SUSY
&  % e+e-
& % pp
&  % Events
&  % PS/Had
&  % DM
&  % SI,SD
&  % Ind DM
&  % b->sg
&   % Bs->mumu
&  %Delta_0-
&  % Delta MBs
&  % b->tau nu
& % (g-2)_mu
& $\surd$ % EW
&  %edm
&  & \cite{sabineWeb}  & \\ % fits 
{\tt SPheno}~\cite{SPHENO} & %L
& $\surd$ %spec
& $\surd$ %nu_R
& $\surd$ %RPV
& %NMSSM
& $\surd$ %CPV
& $\surd$ %FV
& $\surd$ %Decays
& %Decay spin
& % ME
& % ME spin
& % sigma_SUSY
& $\surd$ % e+e-
& % pp
&  % Events
&  % PS/Had
&  % DM
&  % SI,SD
&  % Ind DM
& $\surd$ % b->sg
& $\surd$ % Bs->mumu
&  %Delta_0-
& $\surd$ % Delta MBs
& $\surd$ % b->tau nu
& $\surd$ % (g-2)_mu
& $\surd$% EW
& $\surd$ %edm
&  &\cite{sabineWeb}  & \\ % fits 
{\tt SUSPECT}~\cite{SUSPECT} & %L
& $\surd$ %spec
&  %nu_R
& %RPV
&  %NMSSM
& %CPV
&  %FV
&  %Decays
& %Decay spin
& % ME
& % ME spin
& % sigma_SUSY
&  % e+e-
& % pp
&  % Events
&  % PS/Had
&  % DM
&  % SI,SD
&  % Ind DM
&  $\surd$ % b->sg
&   % Bs->mumu
&  %Delta_0-
&  % Delta MBs
&  % b->tau nu
& $\surd$ % (g-2)_mu
& $\surd$ % EW
&  %edm
&  &\cite{sabineWeb}  & \\ % fits 
\hline
{\tt BRIDGE}~\cite{BRIDGE} & %L
&  %spec
&  %nu_R
& %RPV
&  %NMSSM
& %CPV
&  %FV
& $\surd$ %Decays
& $\surd$ %Decay spin
& % ME
& % ME spin
& % sigma_SUSY
&  % e+e-
& % pp
&  % Events
&  % PS/Had
&  % DM
&  % SI,SD
&  % Ind DM
&  % b->sg
&   % Bs->mumu
&  %Delta_0-
&  % Delta MBs
&  % b->tau nu
& % (g-2)_mu
& % EW
&  %edm
&  &  & \\ % fits 
{\tt CPsuperH}~\cite{CPSUPERH} & %L
& %spec
&  %nu_R
& %RPV
&  %NMSSM
& $\surd$ %CPV
&  %FV
& $\surd$ %Decays
& %Decay spin
& % ME
& % ME spin
& % sigma_SUSY
&  % e+e-
& % pp
&  % Events
&  % PS/Had
&  % DM
&  % SI,SD
&  % Ind DM
& $\surd$ % b->sg
& $\surd$ % Bs->mumu
&  %Delta_0-
& $\surd$ % Delta MBs
& $\surd$ % b->tau nu
& % (g-2)_mu
& % EW
& $\surd$ %edm
&  &  & \\ % fits 
{\tt FCHDECAY}~\cite{FCHDECAY} & %L
&  %spec
&  %nu_R
& %RPV
&  %NMSSM
& %CPV
& $\surd$ %FV
& $\surd$ %Decays
& %Decay spin
& % ME
& % ME spin
& % sigma_SUSY
&  % e+e-
& % pp
&  % Events
&  % PS/Had
&  % DM
&  % SI,SD
&  % Ind DM
&  % b->sg
&   % Bs->mumu
&  %Delta_0-
&  % Delta MBs
&  % b->tau nu
& % (g-2)_mu
& % EW
&  %edm
&  &  & \\ % fits 
{\tt FeynHiggs}~\cite{FEYNHIGGS} & %L
& %spec
&  %nu_R
& %RPV
&  %NMSSM
& $\surd$ %CPV
& $\surd$ %FV
& $\surd$ %Decays
& %Decay spin
& % ME
& % ME spin
& $\surd$ % sigma_SUSY
&  % e+e-
& % pp
&  % Events
&  % PS/Had
&  % DM
&  % SI,SD
&  % Ind DM
&$\surd$  % b->sg
&   % Bs->mumu
&  %Delta_0-
&  % Delta MBs
&  % b->tau nu
& $\surd$ % (g-2)_mu
& $\surd$% EW
&  $\surd$%edm
&  & \cite{FEYNHIGGS}  & \\ % fits 
{\tt HDECAY}~\cite{HDECAY} & %L
&  %spec
&  %nu_R
& %RPV
&  %NMSSM
& %CPV
&  %FV
& $\surd$ %Decays
& %Decay spin
& % ME
& % ME spin
& % sigma_SUSY
&  % e+e-
& % pp
&  % Events
&  % PS/Had
&  % DM
&  % SI,SD
&  % Ind DM
&  % b->sg
&   % Bs->mumu
&  %Delta_0-
&  % Delta MBs
&  % b->tau nu
& % (g-2)_mu
& % EW
&  %edm
&  &  & \\ % fits 
{\tt SDECAY}~\cite{SDECAY} & %L
&  %spec
&  %nu_R
& %RPV
&  %NMSSM
& %CPV
&  %FV
& $\surd$ %Decays
& %Decay spin
& % ME
& % ME spin
& % sigma_SUSY
&  % e+e-
& % pp
&  % Events
&  % PS/Had
&  % DM
&  % SI,SD
&  % Ind DM
&  % b->sg
&   % Bs->mumu
&  %Delta_0-
&  % Delta MBs
&  % b->tau nu
& % (g-2)_mu
& % EW
&  %edm
&  &  & \\ % fits 
\hline
{\tt CalcHEP}~\cite{CALCHEP} &$\supset$ %L
&  %spec
&  %nu_R
& $\surd$ %RPV
& $\surd$  %NMSSM
& $\surd$ %CPV
&  %FV
&  %Decays
& %Decay spin
& $\surd$ % ME
& % ME spin
& $\surd$ % sigma_SUSY
& $\surd$ % e+e-
& $\surd$ % pp
& $\surd$  % Events
&  % PS/Had
&  % DM
&  % SI,SD
&  % Ind DM
&  % b->sg
&   % Bs->mumu
&  %Delta_0-
&  % Delta MBs
&  % b->tau nu
& % (g-2)_mu
& % EW
&  %edm
&  &  & $\surd$ \\ % fits 
{\tt CompHEP}~\cite{COMPHEP} &$\supset$ %L
&  %spec
&  %nu_R
& $\surd$ %RPV
& $\surd$  %NMSSM
& $\surd$ %CPV
&  %FV
&  %Decays
& %Decay spin
& $\surd$ % ME
& % ME spin
& $\surd$ % sigma_SUSY
& $\surd$  % e+e-
& $\surd$ % pp
& $\surd$  % Events
&  % PS/Had
&  % DM
&  % SI,SD
&  % Ind DM
&  % b->sg
&   % Bs->mumu
&  %Delta_0-
&  % Delta MBs
&  % b->tau nu
& % (g-2)_mu
& % EW
&  %edm
&  &  & $\surd$ \\ % fits 
{\tt FormCalc}~\cite{feynarts} & $\surd$ %L
&  %spec
&  %nu_R
& %RPV
&  %NMSSM
& $\surd$ %CPV
& $\surd$ %FV
& %Decays
& %Decay spin
& $\surd$ % ME
& $\surd$ % ME spin
& % sigma_SUSY
& $\surd$ % e+e-
& $\surd$ % pp
&  % Events
&  % PS/Had
&  % DM
&  % SI,SD
&  % Ind DM
&  % b->sg
&   % Bs->mumu
&  %Delta_0-
&  % Delta MBs
&  % b->tau nu
& % (g-2)_mu
& % EW
&  %edm
&  &  & $\surd$ \\ % fits 
{\tt GRACE}~\cite{GRACE} & %L
&  %spec
&  %nu_R
& %RPV
&  %NMSSM
& %CPV
&  %FV
&  %Decays
& %Decay spin
& $\surd$ % ME
& % ME spin
& $\surd$ % sigma_SUSY
&  $\surd$ % e+e-
& % pp
& % Events
&  % PS/Had
&  % DM
&  % SI,SD
&  % Ind DM
&  % b->sg
&   % Bs->mumu
&  %Delta_0-
&  % Delta MBs
&  % b->tau nu
& % (g-2)_mu
& % EW
&  %edm
&  &  & $\surd$ \\ % fits 
{\tt ILCslepton}~\cite{FREITAS} & %L
&  %spec
&  %nu_R
& %RPV
&  %NMSSM
& %CPV
&  %FV
&  %Decays
& %Decay spin
& % ME
& % ME spin
& $\surd$ % sigma_SUSY
& $\surd$  % e+e-
& % pp
&  % Events
&  % PS/Had
&  % DM
&  % SI,SD
&  % Ind DM
&  % b->sg
&   % Bs->mumu
&  %Delta_0-
&  % Delta MBs
&  % b->tau nu
& % (g-2)_mu
& % EW
&  %edm
&  &  & \\ % fits 
{\tt LanHEP}~\cite{LANHEP} & $\surd$ %L
&  %spec
&  %nu_R
& $\surd$ %RPV
& $\surd$ %NMSSM
& $\surd$ %CPV
&  %FV
&  %Decays
& %Decay spin
& % ME
& % ME spin
& % sigma_SUSY
&  % e+e-
& % pp
&  % Events
&  % PS/Had
&  % DM
&  % SI,SD
&  % Ind DM
&  % b->sg
&   % Bs->mumu
&  %Delta_0-
&  % Delta MBs
&  % b->tau nu
& % (g-2)_mu
& % EW
&  %edm
&  &  & \\ % fits 
{\tt MadGraph}~\cite{MADGRAPH} & $\surd$ %L
&  %spec
&  %nu_R
& %RPV
&  %NMSSM
& %CPV
&  %FV
& $\surd$ %Decays
& $\surd$ %Decay spin
&$\surd$ % ME
&$\surd$ % ME spin
&$\surd$ % sigma_SUSY
&$\surd$  % e+e-
& $\surd$ % pp
& $\surd$ % Events
&  % PS/Had
&  % DM
&  % SI,SD
&  % Ind DM
&  % b->sg
&   % Bs->mumu
&  %Delta_0-
&  % Delta MBs
&  % b->tau nu
& % (g-2)_mu
& % EW
&  %edm
&  & \cite{MADGRAPH}  & $\surd$ \\ % fits 
{\tt SUSYGEN}~\cite{SUSYGEN} & %L
&  %spec
&  %nu_R
&$\surd$ %RPV
&  %NMSSM
& $\surd$%CPV
&  %FV
&  %Decays
& %Decay spin
& $\surd$% ME
& $\surd$% ME spin
& % sigma_SUSY
& $\surd$ % e+e-
& % pp
&$\surd$ % Events
&  % PS/Had
&  % DM
&  % SI,SD
&  % Ind DM
&  % b->sg
&   % Bs->mumu
&  %Delta_0-
&  % Delta MBs
&  % b->tau nu
& % (g-2)_mu
& % EW
&  %edm
&  &  & \\ % fits 
{\tt Whizard}~\cite{WHIZARD} &  $\surd$ %L
&  %spec
&  %nu_R
& %RPV
&  %NMSSM
& %CPV
&  %FV
&  %Decays
& %Decay spin
&  $\surd$% ME
&  $\surd$% ME spin
& % sigma_SUSY
&  $\surd$ % e+e-
&  $\surd$ % pp
&   $\surd$% Events
&  % PS/Had
&  % DM
&  % SI,SD
&  % Ind DM
&  % b->sg
&   % Bs->mumu
&  %Delta_0-
&  % Delta MBs
&  % b->tau nu
& % (g-2)_mu
& % EW
&  %edm
&  &  & \\ % fits 
\hline
{\tt Herwig++}~\cite{HERWIG} & $\surd$ %L
&  %spec
&  %nu_R
& $\surd$%RPV
&  %NMSSM
& %CPV
&  %FV
& $\surd$ %Decays
& $\surd$ %Decay spin
& % ME
& % ME spin
& $\surd$% sigma_SUSY
& $\surd$ % e+e-
& $\surd$% pp
& $\surd$ % Events
& $\surd$ % PS/Had
&  % DM
&  % SI,SD
&  % Ind DM
&  % b->sg
&   % Bs->mumu
&  %Delta_0-
&  % Delta MBs
&  % b->tau nu
& % (g-2)_mu
& % EW
&  %edm
&  &  & \\ % fits 
{\tt ISATOOLS}~\cite{ISAJET} & %L
& $\surd$ %spec
& $\surd$ %nu_R
& $\surd$ %RPV
& %NMSSM
& %CPV
&  %FV
& $\surd$ %Decays
& %Decay spin
& % ME
& $\surd$ % ME spin
& $\surd$ % sigma_SUSY
& $\surd$ % e+e-
& $\surd$ % pp
&  $\surd$ % Events
&  $\surd$ % PS/Had
& $\surd$ % DM
&  $\surd$ % SI,SD
&  % Ind DM
& $\surd$ % b->sg
&  $\surd$ % Bs->mumu
&  %Delta_0-
& % Delta MBs
& $\surd$ % b->tau nu
&  $\surd$ % (g-2)_mu
& % EW
&   %edm
&  & \cite{sabineWeb} & \\ % fits 
{\tt PYTHIA}~\cite{PYTHIA} & %L
&  %spec
&  %nu_R
& $\surd$ %RPV
& $\surd$ %NMSSM
& %CPV
&  %FV
& $\surd$ %Decays
& %Decay spin
& % ME
& % ME spin
& $\surd$% sigma_SUSY
& $\surd$ % e+e-
& $\surd$% pp
& $\surd$ % Events
& $\surd$ % PS/Had
&  % DM
&  % SI,SD
&  % Ind DM
&  % b->sg
&   % Bs->mumu
&  %Delta_0-
&  % Delta MBs
&  % b->tau nu
& % (g-2)_mu
& % EW
&  %edm
&  &  & \\ % fits 
{\tt SHERPA}~\cite{SHERPA} &$\surd$ %L
&  %spec
&  %nu_R
& %RPV
&  %NMSSM
& $\surd$ %CPV
& $\surd$ %FV
&  %Decays
& $\surd$ %Decay spin
& $\surd$ % ME
& $\surd$ % ME spin
& $\surd$ % sigma_SUSY
& $\surd$ % e+e-
& $\surd$% pp
& $\surd$ % Events
& $\surd$ % PS/Had
&  % DM
&  % SI,SD
&  % Ind DM
&  % b->sg
&   % Bs->mumu
&  %Delta_0-
&  % Delta MBs
&  % b->tau nu
& % (g-2)_mu
& % EW
&  %edm
&  &  & \\ % fits 
\hline
{\tt HadCalc}~\cite{HADCALC} &$\supset$ %L
&  %spec
&  %nu_R
& %RPV
&  %NMSSM
& $\supset$%CPV
& $\supset$ %FV
&  %Decays
&  %Decay spin
& $\supset$ % ME
& $\supset$% ME spin
& % sigma_SUSY
& $\supset$ % e+e-
& $\supset$% pp
&  % Events
&  % PS/Had
&  % DM
&  % SI,SD
&  % Ind DM
&  % b->sg
&   % Bs->mumu
&  %Delta_0-
&  % Delta MBs
&  % b->tau nu
& % (g-2)_mu
& % EW
&  %edm
&  &  & $\supset$ \\ % fits 
{\tt DarkSUSY}~\cite{DARKSUSY} & %L
& $\surd$ %spec
&  %nu_R
& %RPV
&  %NMSSM
& %CPV
&  %FV
&  %Decays
& %Decay spin
& % ME
& % ME spin
& % sigma_SUSY
&  % e+e-
& % pp
&  % Events
&  % PS/Had
& $\surd$ % DM
& $\surd$ % SI,SD
& $\surd$ % Ind DM
& $\surd$ % b->sg
&   % Bs->mumu
&  %Delta_0-
&  % Delta MBs
&  % b->tau nu
& $\surd$% (g-2)_mu
& % EW
&  %edm
&  & \cite{DARKSUSY}  & \\ % fits 
{\tt micrOMEGAs}~\cite{MICROMEGAS} & $\supset$ %L
& $\supset$ %spec
&  %nu_R
& %RPV
& $\supset$ %NMSSM
& $\surd$%CPV
&  %FV
& $\surd$ %Decays
& %Decay spin
& % ME
& % ME spin
& $\surd$ % sigma_SUSY
& $\surd$ % e+e-
& % pp
&  % Events
&  % PS/Had
& $\surd$ % DM
& $\surd$ % SI,SD
& $\surd$ % Ind DM
& $\surd$ % b->sg
& $\surd$  % Bs->mumu
&  %Delta_0-
& $\supset$ % Delta MBs
& $\supset$ % b->tau nu
& $\surd$% (g-2)_mu
& $\surd$% EW
& $\surd$ %edm
&  & \cite{sabineWeb}  & \\ % fits 
{\tt PROSPINO}~\cite{PROSPINO} & %L
&  %spec
&  %nu_R
& %RPV
&  %NMSSM
& %CPV
& $\surd$ %FV
&  %Decays
& %Decay spin
& % ME
& % ME spin
& $\surd$% sigma_SUSY
& % e+e-
& $\surd$% pp
&  % Events
&  % PS/Had
&  % DM
&  % SI,SD
&  % Ind DM
&  % b->sg
&   % Bs->mumu
&  %Delta_0-
&  % Delta MBs
&  % b->tau nu
& % (g-2)_mu
& % EW
&  %edm
&  &  & \\ % fits 
\hline
{\tt SuperIso}~\cite{SUPERISO} & %L
&  %spec
&  %nu_R
& %RPV
&  %NMSSM
& %CPV
& $\surd$ %FV
&  %Decays
& %Decay spin
& % ME
& % ME spin
& % sigma_SUSY
&  % e+e-
& % pp
&  % Events
&  % PS/Had
&  % DM
&  % SI,SD
&  % Ind DM
&  $\surd$ % b->sg
&  % Bs->mumu
&  $\surd$ %Delta_0-
&  % Delta MBs
&  % b->tau nu
& $\surd$ % (g-2)_mu
& % EW
&  %edm
&  &  & \\ % fits 
{\tt SusyBSG}~\cite{SUSYBSG} & %L
&  %spec
&  %nu_R
& %RPV
&  %NMSSM
& %CPV
&  %FV
&  %Decays
& %Decay spin
& % ME
& % ME spin
& % sigma_SUSY
&  % e+e-
& % pp
&  % Events
&  % PS/Had
&  % DM
&  % SI,SD
&  % Ind DM
&$\surd$  % b->sg
&   % Bs->mumu
&  %Delta_0-
&  % Delta MBs
&  % b->tau nu
& % (g-2)_mu
& % EW
&  %edm
&  &  & \\ % fits 
\hline
{\tt FITTINO}~\cite{FITTINO} & %L
& $\supset$ %spec
&  %nu_R
& %RPV
&  %NMSSM
& %CPV
&  %FV
& $\supset$ %Decays
& %Decay spin
& % ME
& % ME spin
& $\supset$% sigma_SUSY
& $\supset$ % e+e-
& $\supset$ % pp
&  % Events
&  % PS/Had
&  % DM
&  % SI,SD
&  % Ind DM
&  % b->sg
&   % Bs->mumu
&  %Delta_0-
&  % Delta MBs
&  % b->tau nu
& % (g-2)_mu
& % EW
&  %edm
& $\surd$ &  & \\ % fits 
{\tt KISMET}~\cite{Allanach} & %L
& $\supset$ %spec
&  %nu_R
& %RPV
&  %NMSSM
& %CPV
&  %FV
& %Decays
& %Decay spin
& % ME
& % ME spin
& % sigma_SUSY
&  % e+e-
& % pp
&  % Events
&  % PS/Had
& $\supset$ % DM
&  % SI,SD
&  % Ind DM
& $\supset$ % b->sg
& $\supset$  % Bs->mumu
&  %Delta_0-
&  % Delta MBs
&  % b->tau nu
& $\supset$% (g-2)_mu
& $\supset$% EW
&  %edm
& $\surd$ & \cite{Allanach}  & \\ % fits 
{\tt SuperBayeS}~\cite{SUPERBAYES} & %L
& $\supset$ %spec
&  %nu_R
& %RPV
&  %NMSSM
& %CPV
&  %FV
& $\supset$ %Decays
& %Decay spin
& % ME
& % ME spin
& % sigma_SUSY
&  % e+e-
& % pp
&  % Events
&  % PS/Had
& $\supset$ % DM
& $\supset$ % SI,SD
& $\supset$ % Ind DM
& $\surd$ % b->sg
&   % Bs->mumu
&  %Delta_0-
&  % Delta MBs
&  % b->tau nu
& $\supset$% (g-2)_mu
& $\supset$% EW
&  %edm
& $\surd$ & \cite{SPLOTTING}  & \\ % fits 
\hline \end{tabular}
\caption{Summary of functionality of current, publicly available, supported SUSY
  tools. A $\surd$ indicates that there is some support for the feature in
  question, but makes no claims about the accuracy of the calculation. 
$\supset$ indicates that the one of the other packages in the table is
  included in the distribution in order to provide the relevant functionality.
See   section~\protect\ref{sec:summary} for a description of the various
  features. \protect\label{tab:sum}}
\end{sidewaystable}
${\mathcal L}$ indicates that the tool includes or uses a method of encoding a
Lagrangian in order to define extended or new models. 
In the table, `Spectrum' indicates that 
the tool includes an SUSY spectrum calculator, $\nu_R$ indicates that RGEs
include an option for including right-handed neutrinos (and therefore neutrino
mass models), RPV indicates that the tool can handle $R-$parity violation,
NMSSM that it can calculate in the Next-to-Minimal Supersymmetric Standard
Model, CPV that the tool can take into account complex phases in the SUSY
sector and FV that the tool includes some non-minimal flavour violating
effects. `Decays' indicates that the tool automatically calculates the
branching ratios of SUSY or SUSY Higgs decays in the MSSM or
extensions. Tools which have a positive 
entry under `Decay spin' include angular correlation effects from sparticle
spins when simulating decays down cascade decay chains. `ME' indicates a
matrix element generator: the 
code can simulate scattering for 2$\rightarrow N$ hard particles, where $N>2$.
`Initial pol' shows that polarisations of the colliding particles can be taken
into account: usually in $e^+ e^-$ collisions, but sometimes also in $\gamma
\gamma$ or $e \gamma$ collisions. A tick under the $\sigma_{SUSY}$ heading
means that  
the code has an easy user interface for calculating total or differential
cross-sections for the production of (sometimes specified) sparticles and/or
SUSY Higgs. 
$e^+ e^-$ and $pp$ indicates that the initial colliding
particles can be leptonic or hadronic, respectively. `Events' mean that
individual events are simulated, PS/Had that the program can perform parton
showering and/or hadronisation of partons. 
A tick under the $\Omega_{DM} h^2$ header means that the relic density of dark
matter can be calculated, $\sigma_{SI,SD}^p$ that an estimate of dark matter
direct detection is included and `Ind. DM' that some indirect dark matter
detection fluxes are provided. For $b$-observables, $b\rightarrow s \gamma$,
$B \rightarrow \tau \nu, \tau \tau$ and
$B_s \rightarrow \mu^+ \mu^-$ 
indicates that there is a calculation of the relevant branching ratio
including some SUSY effects. A positive entry for $\Delta_{0-}$ means that the
program calculates 
the isospin asymmetry in $B$ decays, whereas an entry under $\Delta M_{B_s}$
that the SUSY 
contributions to $B_{s,d}^0-{\bar B}_{s,d}^0$ mixing are calculated. A
$(g-2)_\mu$ 
entry indicates that a SUSY contribution to the anomalous magnetic moment of
the muon can be easily extracted from the tool, whereas
EW means that
some electroweak observables are provided: usually $\Delta \rho$ and $M_W$.
A tick under edm means that electric dipole moments can be calculated, whereas `Fits' indicates a fitting tool that can fit either collider
observables and/or indirect constraints such as EW observables and dark matter
relic densities. An entry under `Web form' gives the reference
including a link to a web-form where results from the program can be
automatically obtained by filling in a form on the world-wide web.
Finally the `code' column indicates that the package can output computer code,
which can then be compiled into a numerical program in order to evaluate
observables. 
Prospective users are warned
that multi-functionality does not necessarily mean a more accurate
calculation and indeed in some cases, the converse will apply. 
It is hoped that Table~\ref{tab:sum} will help point prospective new users
towards the SUSY tool(s) that they require. 

\acknowledgments
This review has been partially supported by the STFC\@. 
A much shortened form was originally conceived for a talk in the SUSY 07
conference. It borrows heavily from
the manuals and web-pages of the tools discussed, the 
BSM tools repository~\cite{repos} and the SUSY Les Houches Accord
web-pages~\cite{SLHA}. We thank G B\'{e}langer, F Boudjema, T Hahn, S
Heinemeyer, J S Lee, N Mahmoudi, F Maltoni, T Plehn, W Porod, P Richardson, S
Schumann, P 
Slavich and B Webber for corrections and suggestions on the draft.


\begin{thebibliography}{10}
\bibitem{SLHA}
P. Skands, B.C.~Allanach {\em et al}, {\em The SUSY Les Houches Accord:
  Interfacing SUSY Spectrum Calculators, Decay Packages and Event Generators},
JHEP {\bf 0407} (2004) 036, [arXiv:hep-ph/0311123].\\
 \href{http://home.fnal.gov/~skands/slha/}{http://home.fnal.gov/~skands/slha/}

\bibitem{parser}
T.~Hahn,
{\em SUSY Les Houches Accord I/O made easy},
  arXiv:hep-ph/0408283;
T.~Hahn,
  {\em SUSY Les Houches Accord 2 I/O made easy},
  arXiv:hep-ph/0605049.\\
  %%CITATION = HEP-PH/0605049;%%
\href{http://www.feynarts.de/slha/}{http://www.feynarts.de/slha/}


\bibitem{SLHA2}
B.C.~Allanach {\em et al}, 
{\em Susy Les Houches Accord 2}, [arXiv:0801.0045];
{\em ibid}, {\em The SUSY Les Houches Accord Conventions}, in {\em Physics
  Beyond the Standard Model: Supersymmetry}, proceedings of the 2007 Les
Houches ``Physics at TeV Colliders'' Workshop, [{arXiv:0802.3672}].\\
\href{http://home.fnal.gov/~skands/slha/}{http://home.fnal.gov/~skands/slha/}

\bibitem{LHA}
E.~Boos {\em et al}, {\em Generic user process interface for event generators},
in {\em Proceedings of the Workshop on Physics at TeV
  Colliders II Workshop}, [{arXiv:0109068}].

\bibitem{LHA2}
J.~Alwall {\it et al.},
  {\em A standard format for Les Houches event files},
  Comput.\ Phys.\ Commun.\  {\bf 176} (2007) 300
  [arXiv:hep-ph/0609017].
  %%CITATION = CPHCB,176,300;%%

\bibitem{LHA3}
J.~Alwall {\it et al.},
  {\em A Les Houches Interface for BSM Generators},
  arXiv:0712.3311 [hep-ph].
  %%CITATION = ARXIV:0712.3311;%%

\bibitem{Skands:2006sk}
  P.~Z.~Skands,
  {\em A quick guide to SUSY tools},
  arXiv:hep-ph/0601103.
  %%CITATION = HEP-PH/0601103;%%

\bibitem{repos}
P.~Z.~Skands {\it et al.},
 {\em A repository for beyond-the-standard-model tools},
in {\em BSM Working Group summary report of Les Houches at TeV Colliders
  2005}, arXiv:hep-ph/0602198.\\ 
  %%CITATION = FERMILAB-CONF-05-521-T;%%
\href{http://www.ippp.dur.ac.uk/montecarlo/BSM/}{http://www.ippp.dur.ac.uk/montecarlo/BSM/}

\bibitem{ISAJET}
F.~E.~Paige, S.~D.~Protopopescu, H.~Baer and X.~Tata,
  {\em ISAJET 7.69: A Monte Carlo event generator for p p, anti-p p, and e+ e-
  reactions}, 
  arXiv:hep-ph/0312045.\\
  \href{http://www.hep.fsu.edu/~isajet/}{http://www.hep.fsu.edu/~isajet/}
  %%CITATION = HEP-PH/0312045;%%

\bibitem{unofficialISAJET}
C.~Balazs, see {\tt ISALHA.F}, {\tt ISALHD.F} and {\tt LHAISA.F} links on
the SLHA page\\
\href{http://home.fnal.gov/~skands/slha/}{http://home.fnal.gov/~skands/slha/}.

\bibitem{SOFTSUSY}
B.C. Allanach, {\em SOFTSUSY: A program for calculating supersymmetric
  spectra}, Comput. Phys. Commun. {\bf 143} (2002) 305,
[arXiv:hep-ph/0104145].\\
\href{http://projects.hepforge.org/softsusy/}{http://projects.hepforge.org/softsusy/}.

\bibitem{SPHENO}
W.~Porod,
  {\em SPheno, a program for calculating supersymmetric spectra, SUSY particle
  decays and SUSY particle production at e+ e- colliders},
  Comput.\ Phys.\ Commun.\  {\bf 153} (2003) 275
  [arXiv:hep-ph/0301101].\\
  %%CITATION = CPHCB,153,275;%%
\href{http://ific.uv.es/~porod/SPheno.html}{http://ific.uv.es/~porod/SPheno.html}

\bibitem{SUSPECT}
A.~Djouadi, J.~L.~Kneur and G.~Moultaka,
  {\em SuSpect: A Fortran code for the supersymmetric and Higgs particle
    spectrum   in the MSSM},
  Comput.\ Phys.\ Commun.\  {\bf 176} (2007) 426
  [arXiv:hep-ph/0211331]. \\
\href{http://www.lpta.univ-montp2.fr/users/kneur/Suspect/}{http://www.lpta.univ-montp2.fr/users/kneur/Suspect/}

\bibitem{sabineWeb}
S.~Kraml, {\em Comparison of SUSY spectrum generators:
mass spectra, relic densities, etc},\\
\href{http://kraml.home.cern.ch/kraml/comparison/compare.html}{http://kraml.home.cern.ch/kraml/comparison/compare.html}

\bibitem{uncertainties}
B.C.~Allanach, S.~Kraml and W.~Porod,
{\em Theoretical uncertainties in sparticle mass predictions from
  computational tools}, JHEP {\bf 03} (2003) 016, 
{\tt hep-ph/0302102};
B.~C.~Allanach, A.~Djouadi, J.~L.~Kneur, W.~Porod and P.~Slavich,
  {\em Precise determination of the neutral Higgs boson masses in the MSSM},
  JHEP {\bf 0409} (2004) 044
  [arXiv:hep-ph/0406166].
  %%CITATION = JHEPA,0409,044;%%
G.~Belanger, S.~Kraml and A.~Pukhov,
  {\em Comparison of SUSY spectrum calculations and impact on the relic  density
  constraints from WMAP},
  Phys.\ Rev.\  D {\bf 72} (2005) 015003
  [arXiv:hep-ph/0502079].

\bibitem{HERWIG}
 M.~Bahr {\it et al.},
  {\em Herwig++ Physics and Manual},
  arXiv:0803.0883 [hep-ph].\\
  %%CITATION = ARXIV:0803.0883;%%
\href{http://projects.hepforge.org/herwig/}{http://projects.hepforge.org/herwig/}

\bibitem{MADGRAPH}
J.~Alwall {\it et al.},
  {\em MadGraph/MadEvent v4: The New Web Generation},
  JHEP {\bf 0709} (2007) 028
  [arXiv:0706.2334 [hep-ph]];
G.~C.~Cho, K.~Hagiwara, J.~Kanzaki, T.~Plehn, D.~Rainwater and T.~Stelzer,
  {\em Weak boson fusion production of supersymmetric particles at the LHC},
  Phys.\ Rev.\  D {\bf 73} (2006) 054002
  [arXiv:hep-ph/0601063].
  %%CITATION = JHEPA,0709,028;%%
\href{http://cp3wks05.fynu.ucl.ac.be/twiki/bin/view/Main/WebHome}{http://cp3wks05.fynu.ucl.ac.be/twiki/bin/view/Main/WebHome}

\bibitem{PYTHIA}
%T.~Sjostrand, S.~Mrenna and P.~Skands,
%  {\em A Brief Introduction to PYTHIA 8.1},
%  arXiv:0710.3820 [hep-ph];
  %%CITATION = ARXIV:0710.3820;%%
T.~Sjostrand, S.~Mrenna and P.~Skands,
  {\em PYTHIA 6.4 physics and manual},
  JHEP {\bf 0605} (2006) 026
  [arXiv:hep-ph/0603175].\\
  %%CITATION = JHEPA,0605,026;%%
\href{http://home.thep.lu.se/~torbjorn/Pythia.html}{http://home.thep.lu.se/~torbjorn/Pythia.html}

\bibitem{SDECAY}
M.~Muhlleitner, A.~Djouadi and Y.~Mambrini,
  {\em SDECAY: A Fortran code for the decays of the supersymmetric particles  in
  the MSSM},
  Comput.\ Phys.\ Commun.\  {\bf 168} (2005) 46
  [arXiv:hep-ph/0311167].\\
  %%CITATION = CPHCB,168,46;%
\href{http://lappweb.in2p3.fr/~muehlleitner/SDECAY/}{http://lappweb.in2p3.fr/~muehlleitner/SDECAY/}


\bibitem{Richardson:2001df}
  P.~Richardson,
  {\em Spin correlations in Monte Carlo simulations},
  JHEP {\bf 0111} (2001) 029
  [arXiv:hep-ph/0110108].
  %%CITATION = JHEPA,0111,029;%%


\bibitem{BRIDGE}
P.~Meade and M.~Reece,
  {\em BRIDGE: Branching ratio inquiry / decay generated events},
  arXiv:hep-ph/0703031.\\
  %%CITATION = HEP-PH/0703031;%%
\href{http://www.lepp.cornell.edu/public/theory/BRIDGE/}{http://www.lepp.cornell.edu/public/theory/BRIDGE/}



\bibitem{FEYNHIGGS}
S.~Heinemeyer, W.~Hollik and G.~Weiglein,
  {\em FeynHiggs: A program for the calculation of the masses of the neutral
  CP-even Higgs bosons in the MSSM},
  Comput.\ Phys.\ Commun.\  {\bf 124} (2000) 76
  [arXiv:hep-ph/9812320];
  %%CITATION = CPHCB,124,76;%%
S.~Heinemeyer, W.~Hollik and G.~Weiglein,
  {\em The masses of the neutral CP-even Higgs bosons in the MSSM: Accurate
  analysis at the two-loop level},
  Eur.\ Phys.\ J.\  C {\bf 9} (1999) 343
  [arXiv:hep-ph/9812472];
  %%CITATION = EPHJA,C9,343;%%
G.~Degrassi, S.~Heinemeyer, W.~Hollik, P.~Slavich and G.~Weiglein,
  {\em Towards high-precision predictions for the MSSM Higgs sector},
  Eur.\ Phys.\ J.\  C {\bf 28} (2003) 133
  [arXiv:hep-ph/0212020];
  %%CITATION = EPHJA,C28,133;%%
M.~Frank, T.~Hahn, S.~Heinemeyer, W.~Hollik, H.~Rzehak and G.~Weiglein,
  {\em The Higgs boson masses and mixings of the complex MSSM in the
  Feynman-diagrammatic approach},
  JHEP {\bf 0702} (2007) 047
  [arXiv:hep-ph/0611326].\\
  %%CITATION = JHEPA,0702,047;%%
\href{http://www.feynhiggs.de/}{http://www.feynhiggs.de/}


\bibitem{CPSUPERH}
J.~S.~Lee, M.~Carena, J.~Ellis, A.~Pilaftsis and C.~E.~M.~Wagner,
  {\em CPsuperH2.0: an Improved Computational Tool for Higgs Phenomenology in the
  MSSM with Explicit CP Violation},
  arXiv:0712.2360 [hep-ph];
  %%CITATION = ARXIV:0712.2360;%%
  J.~S.~Lee, A.~Pilaftsis, 
  M.~S.~Carena, S.~Y.~Choi, M.~Drees, J.~R.~Ellis and C.~E.~M.~Wagner,
  {\em CPsuperH: A computational tool for Higgs phenomenology in the minimal
  supersymmetric standard model with explicit CP violation},
  Comput.\ Phys.\ Commun.\  {\bf 156} (2004) 283
  [arXiv:hep-ph/0307377].\\
  %%CITATION = CPHCB,156,283;%%
\href{http://www.hep.man.ac.uk/u/jslee/CPsuperH.html}{http://www.hep.man.ac.uk/u/jslee/CPsuperH.html}

\bibitem{HDECAY}
A.~Djouadi, J.~Kalinowski and M.~Spira,
  {\em HDECAY: A program for Higgs boson decays in the standard model and its
  supersymmetric extension},
  Comput.\ Phys.\ Commun.\  {\bf 108} (1998) 56
  [arXiv:hep-ph/9704448].\\
  %%CITATION = CPHCB,108,56;%%
 \href{http://people.web.psi.ch/spira/hdecay/}{http://people.web.psi.ch/spira/hdecay/}

\bibitem{FCHDECAY}
S. B\'{e}jar and J. Guasch.\\
\href{http://fchdecay.googlepages.com/}{http://fchdecay.googlepages.com/}

\bibitem{NMSSMTOOLS}
U.~Ellwanger and C.~Hugonie,
  {\em NMSPEC: A Fortran code for the sparticle and Higgs masses in the NMSSM with
  GUT scale boundary conditions},
  Comput.\ Phys.\ Commun.\  {\bf 177} (2007) 399
  [arXiv:hep-ph/0612134];
  %%CITATION = CPHCB,177,399;%%
U.~Ellwanger, J.~F.~Gunion and C.~Hugonie,
  {\em NMHDECAY: A Fortran code for the Higgs masses, couplings and decay  widths
  in the NMSSM},
  JHEP {\bf 0502} (2005) 066
  [arXiv:hep-ph/0406215].\\
  %%CITATION = JHEPA,0502,066;%%
\href{http://www.th.u-psud.fr/NMHDECAY/nmssmtools.html}{http://www.th.u-psud.fr/NMHDECAY/nmssmtools.html}


\bibitem{feynarts}
T.~Hahn and M.~Perez-Victoria,
  {\em Automatized one-loop calculations in four and D dimensions},
  Comput.\ Phys.\ Commun.\  {\bf 118} (1999) 153
  [arXiv:hep-ph/9807565];
  %%CITATION = CPHCB,118,153;%%
T.~Hahn,
  {\em Generating Feynman diagrams and amplitudes with FeynArts 3},
  Comput.\ Phys.\ Commun.\  {\bf 140} (2001) 418
  [arXiv:hep-ph/0012260];
  %%CITATION = CPHCB,140,418;%%
T.~Hahn and C.~Schappacher,
  {\em The implementation of the minimal supersymmetric standard model in
  FeynArts and FormCalc},
  Comput.\ Phys.\ Commun.\  {\bf 143} (2002) 54
  [arXiv:hep-ph/0105349].\\
  %%CITATION = CPHCB,143,54;%%
\href{http://www.feynarts.de/}{http://www.feynarts.de/}

\bibitem{HADCALC}
M.~Rauch,
{\em Quantum Effects in Higgs-Boson Production Processes at Hadron Colliders},
arXiv:0804.2428 [hep-ph].
%%CITATION = ARXIV:0804.2428;%%
\href{http://www.ph.ed.ac.uk/~mrauch/HadCalc/}{http://www.ph.ed.ac.uk/~mrauch/HadCalc/}


\bibitem{CALCHEP}
A.~Pukhov {\it et al.},
  {\em CompHEP: A package for evaluation of Feynman diagrams and integration  over
  multi-particle phase space. User's manual for version 33},
  arXiv:hep-ph/9908288;
  %%CITATION = HEP-PH/9908288;%%
  A. Pukhov, 
  {\em CalcHEP 3.2: MSSM, structure functions, event generation, batchs, and
  generation of matrix elements for other packages},
  arXiv:hep-ph/0412191.\\
  %%CITATION = HEP-PH/0412191;%%
  \href{http://www.ifh.de/~pukhov/calchep.html}{http://www.ifh.de/~pukhov/calchep.html}

\bibitem{COMPHEP}
E.~Boos {\it et al.}  [CompHEP Collaboration],
  {\em CompHEP 4.4: Automatic computations from Lagrangians to events},
  Nucl.\ Instrum.\ Meth.\  A {\bf 534} (2004) 250
  [arXiv:hep-ph/0403113] and Ref.~\cite{CALCHEP}.
  %%CITATION = NUIMA,A534,250;%%
\href{http://comphep.sinp.msu.ru/}{http://comphep.sinp.msu.ru/}

\bibitem{LANHEP}
A.~V.~Semenov,
  {\em LanHEP: A package for automatic generation of Feynman rules in field
  theory. Version 2.0},
  arXiv:hep-ph/0208011.\\
  %%CITATION = HEP-PH/0208011;%%
\href{http://theory.sinp.msu.ru/~semenov/lanhep.html}{http://theory.sinp.msu.ru/~semenov/lanhep.html}

\bibitem{HELAS}
H.~Murayama, I.~Watanabe and K.~Hagiwara,
  {\em Evaluating cross-sections at TeV energy scale by HELAS},
Tsukuba Workshop JLC 1992 265.
\\ \href{http://www.pas.rochester.edu/~rain/smadgraph/HELAS.ps.gz}{http://www.pas.rochester.edu/~rain/smadgraph/HELAS.ps.gz}
  %%CITATION = KEK-91-11;%%

\bibitem{SHERPA}
T.~Gleisberg, S.~Hoche, F.~Krauss, A.~Schalicke, S.~Schumann and J.~C.~Winter,
  {\em SHERPA 1.alpha, a proof-of-concept version},
  JHEP {\bf 0402} (2004) 056
  [arXiv:hep-ph/0311263].\\
  %%CITATION = JHEPA,0402,056;%%
\href{http://www.sherpa-mc.de/}{http://www.sherpa-mc.de/}

\bibitem{AMEGIC}
F.~Krauss, R.~Kuhn and G.~Soff,
  {\em AMEGIC++ 1.0: A matrix element generator in C++},
  JHEP {\bf 0202} (2002) 044
  [arXiv:hep-ph/0109036].
  %%CITATION = JHEPA,0202,044;%%

\bibitem{WHIZARD}
W.~Kilian, T.~Ohl and J.~Reuter,
  {\em WHIZARD: Simulating Multi-Particle Processes at LHC and ILC},
  arXiv:0708.4233 [hep-ph].\\
\href{http://whizard.event-generator.org/}{http://whizard.event-generator.org/}
  %%CITATION = ARXIV:0708.4233;%%

\bibitem{OMEGA}
M.~Moretti, T.~Ohl and J.~Reuter,
  {\em O'Mega: An optimizing matrix element generator},
  arXiv:hep-ph/0102195.
  %%CITATION = HEP-PH/0102195;%%

\bibitem{SUSYGEN}
N.~Ghodbane,
  {\em SUSYGEN3: An event generator for linear colliders},
  arXiv:hep-ph/9909499.\\
  %%CITATION = HEP-PH/99094
\href{http://lyoinfo.in2p3.fr/susygen/susygen3.html}{http://lyoinfo.in2p3.fr/susygen/susygen3.html}

\bibitem{GRACE}
J.~Fujimoto {\it et al.},
  {\em GRACE/SUSY: Automatic generation of tree amplitudes in the minimal
  supersymmetric standard model},
  Comput.\ Phys.\ Commun.\  {\bf 153} (2003) 106
  [arXiv:hep-ph/0208036].\\
  %%CITATION = CPHCB,153,106;%%
\href{http://minami-home.kek.jp/}{http://minami-home.kek.jp/}



\bibitem{comparison2}
K.~Hagiwara {\it et al.},
  {\em Supersymmetry simulations with off-shell effects for LHC and ILC},
  Phys.\ Rev.\  D {\bf 73} (2006) 055005
  [arXiv:hep-ph/0512260].
  %%CITATION = PHRVA,D73,055005;%%

\bibitem{PROSPINO}
W.~Beenakker, R.~Hopker, M.~Spira and P.~M.~Zerwas,
  {\em Squark and gluino production at hadron colliders},
  Nucl.\ Phys.\  B {\bf 492} (1997) 51
  [arXiv:hep-ph/9610490].
  %%CITATION = NUPHA,B492,51;%
\href{http://www.ph.ed.ac.uk/~tplehn/prospino/}{http://www.ph.ed.ac.uk/~tplehn/prospino/}

\bibitem{FREITAS}
A.~Freitas, A.~von Manteuffel and P.~M.~Zerwas,
  {\em Slepton production at e+ e- and e- e- linear colliders},
  Eur.\ Phys.\ J.\  C {\bf 34} (2004) 487
  [arXiv:hep-ph/0310182].\\
  %%CITATION = EPHJA,C34,487
\href{http://theory.fnal.gov/people/freitas/}{http://theory.fnal.gov/people/freitas/}


\bibitem{Dobbs:2004qw}
  M.~A.~Dobbs {\it et al.},
  {\em Les Houches guidebook to Monte Carlo generators for hadron collider
  physics},
  arXiv:hep-ph/0403045.
  %%CITATION = HEP-PH/0403045;%%


\bibitem{peg}
L.~L\"{o}nnbladd,
{\em ThePEG}\/ Reference Manual. \\
\href{http://projects.hepforge.org/thepeg/doxygen/index.html}{http://projects.hepforge.org/thepeg/doxygen/index.html}

\bibitem{CKKW}
S.~Catani, F.~Krauss, R.~Kuhn and B.~R.~Webber,
  {\em QCD matrix elements + parton showers},
  JHEP {\bf 0111} (2001) 063
  [arXiv:hep-ph/0109231].
  %%CITATION = JHEPA,0111,063;%

\bibitem{SherpaCLModel}
J.~C.~Winter, F.~Krauss and G.~Soff,
  {\em A modified cluster-hadronization model},
  Eur.\ Phys.\ J.\  C {\bf 36} (2004) 381
  [arXiv:hep-ph/0311085].
  %%CITATION = EPHJA,C36,381;%%

\bibitem{wmap5}
G.~Hinshaw {\it et al.}  [WMAP Collaboration],
  {\em Five-Year Wilkinson Microwave Anisotropy Probe (WMAP)
  Observations:Data Processing, Sky Maps, \& Basic Results},
  arXiv:0803.0732 [astro-ph].
  %%CITATION = ARXIV:0803.0732;%%

\bibitem{sloops}
F.~Boudjema, A.~Semenov and D.~Temes,
  {\em SUSY dark matter: Loops and precision from particle physics},
  Nucl.\ Phys.\ Proc.\ Suppl.\  {\bf 157} (2006) 172;
  %%CITATION = NUPHZ,157,172;%%
N.~Baro, F.~Boudjema and A.~Semenov,
  {\em Full one-loop corrections to the relic density in the MSSM: A few
  examples},
  Phys.\ Lett.\  B {\bf 660} (2008) 550
  [arXiv:0710.1821 [hep-ph]].
  %%CITATION = PHLTA,B660,550;%%

\bibitem{DARKSUSY}
P.~Gondolo, J.~Edsjo, P.~Ullio, L.~Bergstrom, M.~Schelke and E.~A.~Baltz,
  {\em DarkSUSY: Computing supersymmetric dark matter properties numerically},
  JCAP {\bf 0407} (2004) 008
  [arXiv:astro-ph/0406204].\\
  %%CITATION = JCAPA,0407,008;%%
\href{http://www.physto.se/~edsjo/darksusy}{http://www.physto.se/~edsjo/darksusy}

\bibitem{MICROMEGAS}
G.~Belanger, F.~Boudjema, A.~Pukhov and A.~Semenov,
  {\em Dark matter direct detection rate in a generic model with micrOMEGAs2.1},
  arXiv:0803.2360 [hep-ph];
  %%CITATION = ARXIV:0803.2360;%%
G.~Belanger, F.~Boudjema, A.~Pukhov and A.~Semenov,
  {\em micrOMEGAs2.0: A program to calculate the relic density of dark matter  in
  a generic model},
  Comput.\ Phys.\ Commun.\  {\bf 176} (2007) 367
  [arXiv:hep-ph/0607059].\\
  %%CITATION = CPHCB,176,367;%%
\href{http://wwwlapp.in2p3.fr/lapth/micromegas/}{http://wwwlapp.in2p3.fr/lapth/micromegas/}


\bibitem{SUSYBSG}
G.~Degrassi, P.~Gambino and P.~Slavich,
  {\em SusyBSG: a fortran code for BR[B $\rightarrow$ Xs gamma] in the MSSM
    with Minimal   Flavor Violation},
  arXiv:0712.3265 [hep-ph].\\
  %%CITATION = ARXIV:0712.3265;%
\href{http://slavich.web.cern.ch/slavich/susybsg/home.html}{http://slavich.web.cern.ch/slavich/susybsg/home.html}

\bibitem{SUPERISO}
F.~Mahmoudi,
  {\em SuperIso: A program for calculating the isospin asymmetry of $B \rightarrow K^* \gamma$
  in the MSSM},
  Comput.\ Phys.\ Commun.\  {\bf 178} (2008) 745
  [arXiv:0710.2067 [hep-ph]].\\
  %%CITATION = CPHCB,178,745;%%
\href{http://www.isv.uu.se/~nazila/superiso/}{http://www.isv.uu.se/~nazila/superiso/}

\bibitem{SUPERBAYES}
R.~Ruiz de Austri and R.~Trotta,
{\em SuperBayeS
Supersymmetry Parameters Extraction Routines for Bayesian Statistics}.\\
\href{http://superbayes.org/}{http://superbayes.org/}


\bibitem{MINUIT}
F.~James, {\em MINUIT: Function minimization and error analysis},
CERN Program Library Long Writeup {\bf D506},\\
\href{http://wwwasdoc.web.cern.ch/wwwasdoc/minuit/minmain.html}{http://wwwasdoc.web.cern.ch/wwwasdoc/minuit/minmain.html}

\bibitem{Buchmueller:2007zk}
  O.~Buchmueller {\it et al.},
  {\em Prediction for the Lightest Higgs Boson Mass in the CMSSM using Indirect
  Experimental Constraints},
  Phys.\ Lett.\  B {\bf 657} (2007) 87
  [arXiv:0707.3447 [hep-ph]].
  %%CITATION = PHLTA,B657,87;%%

\bibitem{roberto}
R.~Trotta,
  {\em Bayes in the sky: Bayesian inference and model selection in cosmology},
  arXiv:0803.4089 [astro-ph].
  %%CITATION = ARXIV:0803.4089;%%

\bibitem{COSMOMC}
A.~Lewis and S.~Bridle, 
{\em Cosmological parameters from CMB and other data: a Monte-Carlo approach},
Phys.\ Rev.\ {\bf D66} (2002) 103511 [arXiv:astro-ph/0205436].\\
\href{http://cosmologist.info/cosmomc}{http://cosmologist.info/cosmomc}

\bibitem{metropolis}
N.~Metropolis, A.W.~Rosenbluth, M.N.~Teller and E.~Teller,
{\em Equations of State Calculations by Fast Computing Machines},
Journal of Chemical Physics, {\bf 21} (1953) 1087-1091

\bibitem{gelmanRubin}
A. Gelman and D. Rubin, 
{\em Inference from Iterative Simulation Using Multiple Sequences},
Stat. Sci. {\bf 7} (1992) 457.

\bibitem{Allanach}
B.C.~Allanach, K.~Cranmer, C.G.~Lester and A.M.~Weber,
{\em  Natural priors, CMSSM fits and LHC weather forecasts},
JHEP {\bf 0708} (2007) 023 [arXiv:0705.0487].\\
\href{http://users.hepforge.org/~allanach/benchmarks/kismet.html}{http://users.hepforge.org/~allanach/benchmarks/kismet.html}

\bibitem{SFITTER}
R.~Lafaye, T.~Plehn, M.~Rauch and D.~Zerwas,
  {\em Measuring Supersymmetry},
  arXiv:0709.3985 [hep-ph].
  %%CITATION = ARXIV:0709.3985;%%

\bibitem{bank}
B.C.~Allanach and C.G.~Lester,
{\em Sampling using a `bank' of clues},
arXiv:0705.0486.


\bibitem{FITTINO}
P.~Bechtle, K.~Desch and P.~Wienemann,
  {\em Fittino, a program for determining MSSM parameters from collider
  observables using an iterative method},
  Comput.\ Phys.\ Commun.\  {\bf 174} (2006) 47
  [arXiv:hep-ph/0412012].\\
  %%CITATION = CPHCB,174,47;%%
\href{http://www-flc.desy.de/fittino/}{http://www-flc.desy.de/fittino/}

\bibitem{Feroz:2007kg}
  F.~Feroz and M.~P.~Hobson,
  {\em Multimodal nested sampling: an efficient and robust alternative to MCMC
  methods for astronomical data analysis},
  arXiv:0704.3704 [astro-ph].
  %%CITATION = ARXIV:0704.3704;%%

\bibitem{Roszkowski:2007fd}
  L.~Roszkowski, R.~Ruiz de Austri and R.~Trotta,
  {\em Implications for the Constrained MSSM from a new prediction for b to s
  gamma},
  JHEP {\bf 0707} (2007) 075
  [arXiv:0705.2012 [hep-ph]].
  %%CITATION = JHEPA,0707,075;%%

\bibitem{SPLOTTING}
{\em Dark Matter Network Exclusion Program}.\\
\href{http://pisrv0.pit.physik.uni-tuebingen.de/darkmatter/}{http://pisrv0.pit.physik.uni-tuebingen.de/darkmatter/}

\bibitem{SPA}
J.~A.~Aguilar-Saavedra {\it et al.},
  {\em Supersymmetry parameter analysis: SPA convention and project},
  Eur.\ Phys.\ J.\  C {\bf 46} (2006) 43
  [arXiv:hep-ph/0511344].
  %%CITATION = EPHJA,C46,43;%%



\end{thebibliography}
\end{document}